\documentclass[amsthm]{elsart}

\usepackage{graphicx}
\usepackage{amssymb, amsmath}
\usepackage{bm}
\usepackage{color}
\usepackage{CJK}
\makeatletter
\newif\if@restonecol
\makeatother

\usepackage[ruled,lined,algonl]{algorithm2e}
\usepackage{dcolumn}
\usepackage{amsmath}
\usepackage{cases}
\usepackage{CJK}
\usepackage{yjsco}
\usepackage{natbib}
\usepackage{booktabs}
\IncMargin{2em}  %\incmargin{2em} 
\LinesNumbered %\linesnumbered

\def\note#1{\normalmarginpar\marginpar{\parbox{5cm}{\begin{flushleft}\footnotesize #1\end{flushleft}}}}

\newcommand{\xvar}{\bm{x}}

\DeclareMathOperator{\lt}{ht}
\DeclareMathOperator{\lm}{hm}

\DeclareMathOperator{\cls}{cls}
\DeclareMathOperator{\ldeg}{ldeg}
\DeclareMathOperator{\ini}{ini}
\DeclareMathOperator{\lc}{lc}
\DeclareMathOperator{\lv}{lv}
\DeclareMathOperator{\pquo}{pquo}
\DeclareMathOperator{\prem}{prem}
\DeclareMathOperator{\lcm}{lcm}
\DeclareMathOperator{\res}{res}
\newcommand{\field}[1]{\mathcal{#1}}
\newcommand{\pset}[1]{\mathcal{#1}}
\newcommand{\bases}[1]{\langle #1 \rangle}
\newcommand{\point}[1]{\bm{#1}}

\newcommand{\qnum}{\mathbb{Q}}
\newcommand{\nf}{{\field{K}}}
\newcommand{\kxring}{\field{K}[\point{x}]}

\newcommand{\kx}{\kxring}

\newcommand{\kxonexn}{\nf[x_1,\ldots,x_n]}
\newcommand{\gb}{Gr\"{o}bner~}
\newcommand{\grobner}{Gr\"{o}bner}
\newcommand{\chongfen}{$(\Rightarrow)$~}
\newcommand{\biyao}{$(\Leftarrow)$~}

\newcommand{\rli}[1]{}

\DeclareMathOperator{\zero}{Zero}
\DeclareMathOperator{\nops}{nops}

\DeclareMathOperator{\algbasicset}{\sf BasSet}

\DeclareMathOperator{\stprem}{stprem}
\DeclareMathOperator{\algcharset}{\sf NewCharSet}
\DeclareMathOperator{\algcharsetw}{\sf NewCharSetw}
\DeclareMathOperator{\alggencharset}{\sf GenCharSet}
\DeclareMathOperator{\algcs}{\sf CharSet}
\DeclareMathOperator{\algcsw}{\sf CharSetw}
\DeclareMathOperator{\alggb}{\sf Groebner}
\DeclareMathOperator{\algautoset}{\sf MedSet}
\DeclareMathOperator{\algfind}{\sf Find3R}
\DeclareMathOperator{\algrem}{\sf Rem}
\DeclareMathOperator{\algremplus}{\sf RemCh}

\newcommand{\rd}{\mathfrak{D}}
\newcommand{\rdset}{\mathbb{D}}

\renewcommand{\theenumi}{(\alph{enumi})}
\CJKtilde
\begin{document}\begin{CJK*}{GBK}{song}
\newtheorem{theorem}{\hskip \parindent Theorem}[section]
\newtheorem{proposition}[theorem]{\hskip \parindent Proposition}
\newtheorem{lemma}[theorem]{\hskip \parindent Lemma}
\newtheorem{example}[theorem]{\hskip \parindent Example}
%\newtheorem{axiom}[theorem]{\hskip \parindent Axiom}
%\newtheorem{corollary}[theorem]{\hskip \parindent Corollary}
%\theorembodyfont{\rmfamily}
%\newtheorem{definition}[theorem]{\hskip \parindent definition}
%\newtheorem{problem}[theorem]{\hskip \parindent Problem}
%\theoremstyle{nonumberplain}
%\newtheorem{remarks}{\hskip \parindent \hei 注}
%\theoremsymbol{$\square$}
%\theoremseparator{.}
%\newtheorem{pf}{\hskip \parindent pf}

\begin{frontmatter}

\title{A New Algorithmic Scheme for Computing Characteristic Sets}

\thanks{This work has been supported partially by the Open Fund of SKLSDE under Grant No.\ SKLSDE-2011KF-02 and
the ANR-NSFC project EXACTA (ANR-09-BLAN-0371-01/60911130369).}

\author{Meng Jin }
\address{LMIB--School of Mathematics and Systems Science, Beihang University, Beijing 100191, China}
\ead{jinmeng101@gmail.com}
%\ead[url]{URL 1}

\author{Xiaoliang Li}
\address{LMIB--School of Mathematics and Systems Science, Beihang University, Beijing 100191, China}
\ead{xiaoliangbuaa@gmail.com}
%\ead[url]{URL 2}
\author{Dongming Wang}
\address{Laboratoire d'Informatique de Paris 6, CNRS -- Universit\'e Pierre et Marie Curie, 4 place Jussieu -- BP
169, 75252 Paris cedex 05, France} \ead{Dongming.Wang@lip6.fr}
%\ead[url]{www-calfor.lip6.fr/\textasciitilde wang/}
\begin{abstract}
Ritt-Wu's algorithm of characteristic sets is the most
representative for triangularizing sets of multivariate polynomials.
Pseudo-division is the main operation used in this algorithm. In
this paper we present a new algorithmic scheme for computing
generalized characteristic sets by introducing other admissible
reductions than pseudo-division. A concrete subalgorithm is designed
to triangularize polynomial sets using selected admissible
reductions and several effective elimination strategies and to
replace the algorithm of basic sets (used in Ritt-Wu's algorithm).
The proposed algorithm has been implemented and experimental results
show that it performs better than Ritt-Wu's algorithm in terms of
computing time and simplicity of output for a number of non-trivial
test examples.
\end{abstract}


\end{frontmatter}

\section{Introduction}

For solving systems of multivariate polynomial equations there are
mainly three elimination approaches based on resultants, triangular
sets, and Gr\"{o}bner bases (see, e.g., \citealt{KL1992E} and
\citealt{w01e}). One of the best known concepts for triangular sets
is characteristic set, which was introduced first by \cite{r50d} for
prime ideals. Since 1980, W.-T. Wu has considerably developed Ritt's
theory and method of characteristic sets by removing irreducibility
requirements and designing efficient algorithms for zero
decomposition of arbitrary polynomial systems. Ritt-Wu's method has
been improved and extended by a number of researchers and has been
successfully applied to many problems in science and engineering
(see, e.g., \citealt{WX2007M}).

To speed up the computation of characteristic sets, \cite{Chou1988M}
and later \cite{CG1990R} introduced the notions of W-prem and
W-characteristic set. Characteristic sets and their complexity were
studied by \cite{GM1991E,GM1991W} and new algorithms for computing
characteristic sets with simple exponential sequential and
polynomial parallel time complexities were presented.
\cite{Wang1992A} proposed an effective strategy to improve Ritt-Wu's
algorithm of characteristic sets. \cite{g92s} presented a method
based on Ritt-Wu's algorithm to deal with parametric polynomial
systems. Certain properties about ascending chains were studied and
used to enhance the efficiency of Ritt-Wu's algorithm in
\cite{g93d}. A complete implementation of Ritt-Wu's method in the
Maple system was reported in \cite{Wang1995A}. \cite{Wang2001A} also
generalized Ritt-Wu's algorithm by means of one-step
pseudo-reduction with strategies for the selection of reductends and
optimal reductors. \cite{Li2006S} described a modified Ritt-Wu
algorithm that can avoid redundant decompositions.

Besides Ritt-Wu's, there are many other efficient methods for
decomposing systems of multivariate polynomials. \cite{k93g}
introduced the notion of regular chain and presented a method for
decomposing any algebraic variety into unmixed-dimensional
components represented by regular chains. Regular chain was also
defined independently by \cite{YZ1994S} under the name of normal
ascending set. \cite{l91n} introduced the concept of normalized
triangular set (which is a special regular chain) and provided a
method for uniquely decomposing the zero set of any polynomial set
into regular zero sets of normalized triangular sets. Another method
of triangular decomposition was proposed by \cite{w93e}. \cite{a99t}
studied the relationship between several different notions of
triangular sets used in these methods and proved the equivalence of
four notions (including regular chain). Comprehensive investigations
on triangular sets and triangularizing algorithms were also carried
out in \cite{Hub2003}. The reader may refer to
\cite{Sza1999C,a99ta,m00t,w00c,w01e,CM2011A} and references therein
for other related work on triangular decomposition of polynomial
systems. Extensions of the theories and methods in the algebraic
case to the differential, difference, and other cases may be found
in {\cite{BLOP1995R,LW1999,Hub2000F,CG2003I,Hub2003N,BLFP2009C,GLC2009A,BLM2010C}}
and references given therein.

Pseudo-division is the main operation used in most of the
above-mentioned methods. In the process of computing a
characteristic set, one needs to construct polynomial remainder
sequences~(PRSs). The coefficients of intermediate polynomials in a
PRS may swell dramatically. As pointed out by \cite{Col1967S}, the
coefficients of polynomials in an Euclidean PRS (computed using
pseudo-division) may grow exponentially. An exponential upper bound
is given in \citet[(27) in 4.6.1]{Knuth1997T}. There are several
other PRSs, such as subresultant PRS in which the coefficients of
polynomials can be determined by using certain matrices (see
\citealt{GL2003S} and \citealt{Mis1993A}) and swell much more slowly
than those in an Euclidean PRS. \cite{Li1989D} was the first who
studied superfluous factors appearing in the computation of
characteristic sets.

In this paper we follow the work of \cite{Wang2001A} to present a
new algorithmic scheme for computing generalized characteristic sets
efficiently. To compute characteristic sets, one can replace
pseudo-division by one-step pseudo-reduction, but this does not
enhance the efficiency very much because extraneous factors for the
pseudo-rests may be created (see the analysis in \citealt{SAS1984T}
and \citealt{Wang2001A}). Our intention here is to design a
reduction mechanism that can take advantage of the structure and
properties of the polynomials under consideration and thus provide
more flexibility for reduction strategies. A polynomial set which
generates the same ideal as the input polynomial set is maintained,
leading to the concept of generalized characteristic set. This set
may contain polynomials of degrees smaller than the degrees of those
in the input set and therefore may take less computing time for
pseudo-reduction to $0$. By introducing admissible reductions other
than pseudo-division, it is possible to control the swell of
coefficients of intermediate polynomials and to compute with smaller
polynomials. Several strategies for concrete admissible reductions
are adopted in a sample algorithm of the new scheme. Two versions of
the sample algorithm have been implemented and compared with three
algorithms implemented in Maple: two in the Epsilon package and one
in the Groebner package. More details about the outputs of these
algorithms for some examples are given in the appendix.

The paper is structured as follows. Section 2 consists mainly of
terminologies and notations about characteristic sets. In Section 3,
we describe the main {algorithmic scheme} for computing generalized
characteristic sets. A concrete subalgorithm is designed to
triangularize polynomial sets by means of admissible reductions and
effective elimination strategies and to replace the algorithm of
basic sets (used in Ritt-Wu's algorithm). The termination and
correctness of these algorithms are proved. In Section 4,
discussions on concrete admissible reductions are given. Section 5
provides details for the sample subalgorithm. Section 6 contains an
illustrative example and some experimental results, showing that our
proposed algorithm performs better than Ritt-Wu's in terms of
efficiency and simplicity of output for a number of non-trivial test
examples.

\section{Preliminaries}\label{sec:prelim}

Let $\nf$ be any field. The notation $\kx$ or $\kxonexn$ stands for
the ring of polynomials over $\nf$ in the variables
{$x_1,\ldots,x_n$}. In what follows, we always assume that the
variables are ordered as $x_1 <\cdots < x_n$.

Let $F$ be an arbitrary non-zero polynomial in $\kx$. We use
$\deg(F,x_k)$ and $\lc(F,x_k)$ to denote the \emph{degree} and the
\emph{leading coefficient} of $F$ with respect to (w.r.t.) the
variable $x_k$ respectively. The biggest variable that effectively
appears in $F$ is {called} the \emph{leading variable} of $F$ and
denoted by $\lv(F)$. Moveover, $\ini(F):=\lc(F, \lv(F))$ and
$\ldeg(F):=\deg(F, \lv(F))$ are called the \emph{initial} and the
\emph{leading degree} of $F$ respectively. For any non-constant
polynomial $F$, the index $c$ of its leading variable $x_c$ is
called the \emph{class} of $F$ and denoted by $\cls(F)$. The class
of any non-zero constant is set to be $0$.

\begin{defn}
    An ordered polynomial set $[T_1,\ldots,T_r]\subseteq\kx\setminus\nf$
    is called a \emph{triangular set} in $\kx$ if $\lv(T_1)<\cdots<\lv(T_r)$.
\end{defn}

\begin{defn}
Let {$P$ and $Q$} be two polynomials in $\kx$. We say that $P$ is
\emph{reduced} w.r.t.\ $Q$ if $\deg(P,\lv(Q))<\ldeg(Q)$; otherwise,
$P$ is \emph{reducible} w.r.t.\ $Q$.

Furthermore, let $\pset{T}=[T_1,\ldots,T_r]$ be a triangular set in
$\kx$. $F$ is  \emph{reduced} w.r.t.\ $\pset{T}$ if for every
$i=1,\ldots,r$, $F$ is reduced w.r.t.\ $T_i$; {otherwise, $F$ is
\emph{reducible} w.r.t.\ $\pset{T}$}. The triangular set $\pset{T}$
is called  an \emph{auto-reduced} (or \emph{initial-reduced}) set if
for every $i=2,\ldots,r$, $T_i$ (or $\ini(T_i)$) is reduced w.r.t.\
$[T_1,\ldots,T_{i-1}]$.
\end{defn}

\begin{defn}\label{def:ranking}
Let $P$ and $Q$ be non-zero polynomials in $\kx$. $P$ is said to
have \emph{a lower ranking} than $Q$, denoted as {$P\prec Q$}, if {$\cls(P)<\cls(Q)$, or $\cls(P)=\cls(Q)$ and
$\ldeg(P)<\ldeg(Q)$.} In this case, we also say that $Q$ has \emph{a higher ranking} than $P$ {and denote it as
$Q\succ P$}.

If neither $P\prec Q$ nor $Q\succ P$ holds, then we say that $P$ and
$Q$ \emph{have the same ranking} and write $P\thicksim Q$. Moreover,
$P\succ Q$ or $P\thicksim Q$ is denoted as $P\succsim Q$, and
$P\precsim Q$ can be similarly defined.
\end{defn}

\begin{defn}
Let $\pset{T}=[T_1,\ldots,T_{r}]$ and $\pset{S}=[S_1,\ldots,S_s]$ be two triangular sets in $\kx$. $\pset{T}$ is said to have \emph{a lower ranking} than $\pset{S}$, denoted as {$\pset{T}\prec \pset{S}$}, if
\begin{enumerate}
   \item there exists an $i\leq \min({r},s)$ such that $T_1\thicksim S_1,\ldots,T_{i-1}\thicksim S_{i-1}$ and $T_i\prec S_i$; or
   \item ${r}>s$ and $T_1\thicksim S_1,\ldots,T_{s}\thicksim S_{s}$.
\end{enumerate}
In this case, we also say that $\pset{S}$ has \emph{a higher ranking} than $\pset{T}${ and denote it as $\pset{S}\succ \pset{T}$}.

If neither $\pset{T}\succ\pset{S}$ nor $\pset{T}\prec\pset{S}$, {then} $\pset{T}$ and $\pset{S}$ are said to
 have \emph{the same ranking} and denoted as $\pset{T}\thicksim\pset{S}$. In this case, one can easily
 know that ${r}=s$ and $T_1\thicksim S_1,\ldots,T_{r}\thicksim S_{r}$.
\end{defn}

The order $\succsim$ defined above is a partial order and can be
used to order triangular sets. The following proposition indicates
that any {non-empty} set of triangular sets has a \emph{minimal
element}.

\begin{prop}[\text{\citealt{w86b}}]
\label{prop:zhi-finite} For any triangular set sequence
$\pset{T}_1\precsim\pset{T}_2\precsim\pset{T}_3\precsim\cdots$,
there exists an integer {$m$} such that for any $i\geq m$,
$\pset{T}_i\,{\thicksim}\,\pset{T}_m$.
\end{prop}

An {(initial-reduced)} auto-reduced  set in $\kx$ is also
called a (\emph{weak}) \emph{non-contradictory ascending set}, while
$[a]~(a\in\nf\setminus\{0\})$ is called a (\emph{weak})
\emph{contradictory ascending set}. Both of them are jointly called
(\emph{weak}) \emph{ascending sets}.

Let $\pset{P}$ be a non-empty polynomial set in $\kx$ and $\Omega$
denote the set of all the ascending sets that are contained in
$\pset{P}$; $\Omega$ is not empty because the set of any single
polynomial in $\pset{P}$ is an ascending set by definition. By
Proposition \ref{prop:zhi-finite}, $\Omega$ has a minimal element,
which is {called} a \emph{basic set} of $\pset{P}$. This set can be
computed, e.g., by {the algorithm} $\algbasicset$ described in
\cite{w01e}.

\begin{defn}
 Let $\pset{P}$ be a non-empty polynomial set $\kx$. {An} ascending set $\pset{M}$ is called a
 \emph{medial set} of $\pset{P}$, if {$\pset{M}\subseteq\bases{\pset{P}}$
 and $\pset{M}$ has ranking not higher than the ranking of any basic set of $\pset{P}$.}
\end{defn}

\begin{prop}[\text{\citealt{w01e}}]\label{prop:med-prec}
Let $\pset{P}$ be a non-empty polynomial set in $\kx$,
$\pset{M}=[M_1,\ldots,M_r]$ a medial set of $\pset{P}$ with
$M_1\not\in\nf$, and $M$ a non-zero polynomial reduced w.r.t.\
$\pset{M}$. Then for any medial set $\pset{M}'$ of
$\pset{P}'=\pset{P}\cup\pset{M}\cup\{M\}$, we have $\pset{M}'\prec
\pset{M}$.
\end{prop}

For any $F, G\in\kx$ with $G\neq 0$, let $\prem(F,G,x_k)$ and
$\pquo(F,G,x_k)$ denote respectively the pseudo-remainder and the
pseudo-quotient of $F$ w.r.t.\ $G$ in $x_k$. For any triangular set
$\pset{T}=[T_1,\ldots,T_{r}]$ and polynomial set $\pset{P}$ in
$\kx$, define $\prem(F, \pset{T}) :=
\prem(\cdots\prem(F,T_r,\lv(T_r)),\ldots,T_1,\lv(T_1))$ and
$\prem(\pset{P}, \pset{T}) := \{\prem(P,
\pset{T}):\,P\in\pset{P}\}$.

\begin{defn}\label{def:charsetwu}
{A} (weak) ascending set $\pset{C}$ in $\kx$ is called a
(\emph{weak}) \emph{characteristic set} of a non-empty polynomial
set $\pset{P}\subseteq\kx$ if {$\pset{C}\subseteq\bases{\pset{P}}$
and $\prem(\pset{P},\pset{C})=\{0\}$.}
\end{defn}

For characteristic sets, the following property is fundamental.

\begin{prop}[\citealt{w86b}]\label{prop:char-prop}
Let $\pset{C}=[C_1,\ldots,C_r]$ be any characteristic set of a
polynomial set $\pset{P}\subseteq\kx$ and
$$I_i=\ini({C}_i),\quad\pset{P}_i=\pset{P}\cup\{I_i\},\quad \pset{I}={\{I_1,\ldots,I_r\}}.$$
Then
  \begin{equation}\label{eq:zerorelation}
    \begin{split}
        & \zero(\pset{C}/\pset{I})\subseteq \zero(\pset{P}) \subseteq \zero(\pset{C}), \\
         & \zero(\pset{P})=\zero(\pset{C}/\pset{I})\cup\bigcup_{i=1}^r \zero(\pset{P}_i).
     \end{split}
  \end{equation}

\end{prop}

\section{Algorithmic Scheme for Computing Generalized Characteristic Sets}

In this section we generalize the concept of characteristic set and
give a proposition to show that the generalized concept preserves
the basic property (Proposition \ref{prop:char-prop}) of the
original one. For computing generalized characteristic sets, we
propose a new algorithmic scheme, in which the set of reductions,
the termination condition, and strategies for finding reduction
polynomials can be viewed as placeholders. The scheme will be
instantiated as concrete algorithms in Section~\ref{sec:samsub}.

\begin{defn}\label{def:gencharset}
For any non-empty polynomial set $\pset{P}$ in $\kx$, an ascending
set $\pset{C}\subseteq\kx$ is called a \emph{generalized
characteristic set} of $\pset{P}$ if
  \begin{enumerate}
    \item $\pset{C}\subseteq\bases{\pset{P}}$;
    \item there exists a polynomial set $\pset{Q}\subseteq\kx$ such that $\bases{\pset{Q}}=\bases{\pset{P}}$
    and $\prem(\pset{Q},\pset{C})=\{0\}$.
  \end{enumerate}
\end{defn}

It is easy to see that characteristic sets are indeed a special kind
of generalized characteristic sets. The zero relations in
\eqref{eq:zerorelation} also hold for generalized characteristic
sets.

\begin{prop}\label{prop:gch-prop}
Let $\pset{C}=[C_1,\ldots,C_r]$ be any generalized characteristic
set of a polynomial set $\pset{P}\subseteq\kx$ and
  $$I_i=\ini({C}_i),\quad\pset{P}_i=\pset{P}\cup\{I_i\},\quad \pset{I}={\{I_1,\ldots,I_r\}}.$$
Then the zero relations in \eqref{eq:zerorelation} hold.
\end{prop}

\begin{pf}
By Definition \ref{def:gencharset}, there exists a
$\pset{Q}\subseteq\kx$ such that $\bases{\pset{P}}=
\bases{\pset{Q}}$ and $\prem(\pset{Q},\pset{C})=\{0\}$. It is easy
to know that $\pset{C}$ is a characteristic set of $\pset{Q}$.
Bearing in mind Proposition~\ref{prop:char-prop} and
$\zero(\pset{P})=\zero(\pset{Q})$, one sees clearly the truth of the
proposition.
\end{pf}

Now we can describe the main algorithm $\algcharset$ (Algorithm
\ref{alg:charset}), which computes a generalized characteristic set
of any given non-empty polynomial set in $\kx$.

\medskip
\begin{algorithm}[htb]\label{alg:charset}
%\caption{$\pset{A}:=\algcharset(\pset{F},\rdset)$}

\KwIn{$\pset{F}$ --- non-empty polynomial set in $\kx$;\linebreak
$\rdset$ --- set of admissible reductions in $\kx$.}

\KwOut{$\pset{A}$ --- generalized characteristic set of $\pset{F}$.}

\BlankLine

$\pset{G}:=\pset{F}$; $\pset{R}:=\pset{F}$\;

\While{$\pset{R}\neq\emptyset$}{

$[\pset{A},\pset{B}]:=\algautoset(\pset{G},\rdset)$\;\label{ln:autoset}
%\tcp*[h]{O(Left,This)==1}
    \eIf{$\pset{A}$~{ is a contradictory ascending set}}{
        $\pset{R}:=\emptyset$\;
    }{
        $\pset{R}:=\prem(\pset{B}\setminus\pset{A},\pset{A})\setminus\{0\}$\;\label{ln:cs-prem}
        $\pset{G}:=\pset{G}\cup\pset{A}\cup\pset{R}$\;\label{ln:cs-g}
    }
}
\end{algorithm}

Algorithm \ref{alg:charset} is similar to the algorithm
$\alggencharset$ presented in \cite{w01e}. The {difference} is that
$\algautoset$ here outputs not only a medial set $\pset{A}$ of the
polynomial set $\pset{G}$, but also another basis $\pset{B}$ of the
ideal $\bases{\pset{G}}$. Properly storing in $\pset{B}$ the
information from computing $\pset{A}$, one may check more
efficiently whether $\pset{A}$ is a characteristic set, which
usually takes a large proportion of the total computing time. Before
describing the concrete steps of $\algautoset$, we need some
additional definitions such as admissible reduction.

\begin{defn}\label{def:rankofpol}
Let $P,Q\in\kx\setminus\{0\}$ and $<_{\rm lex}$ be the lexicographic
order in $\kx$ such that $x_1<\cdots<x_n$. Denote $P<Q$ or $Q>P$ if
  \begin{enumerate}
    \item $\lt(P)<_{\rm lex}\lt(Q)$, or
    \item $\lt(P)=\lt(Q)$ and $P-\lm(P)<Q-\lm(Q)$,
  \end{enumerate}
  where $\lt(P)$ and $\lm(P)$ stand for the \emph{heading term} and the \emph{heading monomial} of $P$ under $<_{\rm lex}$ respectively.
    Set $0<P$ for any $P\in\kx\setminus\{0\}$.

  If neither $P<Q$ nor $P>Q$, we write $P\thickapprox Q$. The notation $P\gtrapprox Q$ means that
  $P>Q$ or $P\thickapprox Q$, and $P\lessapprox Q$ is similarly defined.
\end{defn}

The partial order $<$ is a refinement of $\prec$ (Definition
\ref{def:ranking}). More precisely, for any
$P,Q\in\kx\setminus\{0\}$, $P\prec Q$ implies that $P<Q$. On the
contrary, $P<Q$ implies only $P\precsim Q$, and $P\prec Q$ does not
necessarily hold. Consider for example the polynomials $P=y^3+x$ and
$Q=y^3+x^2$ in $\nf[x,y]$. By definition, $P<Q$ and $P\thicksim Q$.

\begin{lem}\label{lem:fseq}
For any polynomial sequence $P_1\gtrapprox P_2\gtrapprox P_3\gtrapprox \cdots$,
there exists an integer $m$ such that for any $i\geq m$, $P_i\thickapprox P_m$.
\end{lem}
\begin{pf}
From $P_1\gtrapprox P_2\gtrapprox P_3\gtrapprox \cdots$, one can
obtain the sequence $\lt(P_1)\geq \lt(P_2)\geq \lt(P_3)\geq \cdots$.
In view of the property of term orders (see, e.g., \citealt[Lemma 2
in Chapter 2]{c97i}), there exists an integer $s$ such that for any
$i\geq s$, $\lt(P_i)=\lt(P_s)$.

Let $P'_i=P_i-\lm(P_i)$. We have a new sequence $P'_s\gtrapprox
P'_{s+1}\gtrapprox P'_{s+2}\gtrapprox \cdots$. Similarly, there
exists an integer $t~(\geq s)$ such that for any $i\geq t$,
$\lt(P'_i)=\lt(P'_t)$. The rest can be done in the same manner. As
the number of terms of any polynomial is limited, we will finally
obtain an integer $m$ such that for any $i\geq m$, all $P_i$ have
the same set of terms, which means $P_i\thickapprox P_m$.
\end{pf}

Let $\rd$ be an operation with two {polynomials}
$P,Q\in\kx\setminus\nf$ as its input and {an} ordered set of two
polynomials in $\kx$ as its output, and the latter is denoted by
$\algrem(P,Q,\rd)$. The definition of admissible reduction, which is
somewhat abstract, is given below. In the next section, we will
discuss a number of concrete reductions used in our experiments.

\begin{defn}\label{def:weakred}
Let $[R_1,R_2]=\algrem(P,Q,\rd)$ for any $P,Q\in\kx\setminus\nf$.
The operation $\rd$ is called an \emph{admissible reduction} (or
\emph{reduction} for short if there is no confusion) in $\kx$ if
$R_1,R_2\in\bases{P,Q}$.

Suppose that $\rd$ is an admissible reduction in $\kx$. We say that
$P$ is \emph{$\rd$-reducible} w.r.t.\ $Q$ if $P<R_1$ and
$Q\lessapprox R_2$; otherwise, $P$ is \emph{$\rd$-reduced} w.r.t.\
$Q$. Moreover, $P$ and $Q$ are called the \emph{reductend} and the
\emph{reductor} respectively, and $\algrem(P,Q,\rd)$ the \emph{$\rd$
reduction-rest} of $P$ {by} $Q$.
\end{defn}

The subalgorithm $\algautoset$ is given as Algorithm
\ref{alg:autoset}, where \textsf{cond} can be replaced by any
Boolean expression that guarantees the termination of the
\textbf{while} loop.

The operation $\algfind(\pset{A},\rdset)$ chooses polynomials $P$
(reductend) and $Q$ (reductor) from $\pset{A}$ and $\rd$ (reduction)
from $\rdset$ such that $P$ is $\rd$-reducible w.r.t.\ $Q$. If
$\algfind(\pset{A},\rdset)$ finds appropriate $P,Q,\rd$, then it
returns this triple; otherwise, it returns $[\,]$.

The function $\algremplus$ is an extension of $\algrem$. $\algremplus(P,Q,\rd)$ returns not only the
 $\rd$ reduction-rest $\pset{R}$ of $P$ by $Q$, but also a Boolean value $b$. When $b$ is \textbf{true},
the reduction-rest $\pset{R}$ must satisfy the condition
$P,Q\in\bases{\pset{R}}$. The computation of $b$ depends on the
concrete admissible reduction $\rd$, which will be discussed in the
next section. The \textbf{if} block (lines
\ref{ln:PQG}--\ref{ln:end-if}) is designed to store information
acquired in the reduction process, which may be used by
$\algcharset$ to check more efficiently whether the output
$\pset{M}$ is a characteristic set.

In algorithm $\algautoset$, one may view $\textsf{cond}$, $\rd$ and
$\algfind$ as placeholders, which will be instantiated in Section
\ref{sec:samsub}. The correctness and termination of Algorithms
{\ref{alg:autoset} and \ref{alg:charset}} are proved as follows.

\begin{algorithm}[ht]\label{alg:autoset}%\caption{$\pset{M},\pset{G}:=\algautoset(\pset{F},\rdset)$}

\KwIn{$\pset{F}$ --- non-empty polynomial set in $\kx$;\linebreak
$\rdset$ --- set of admissible reductions in $\kx$.}

\KwOut{$\pset{M}$ --- medial set of $\pset{F}$;\linebreak
$\pset{G}$
--- polynomial set satisfying
$\bases{\pset{G}}=\bases{\pset{F}}$.}

\BlankLine

$\pset{A}:=\pset{F}$; $\pset{G}:=\pset{F}$\;

\While{\textup{\textsf{cond}}}{

$[P,Q,\rd]:=\algfind(\pset{A},\rdset)$\;

    $[\pset{R},b]:=\algremplus(P,Q,\rd)$\;\label{ln:rem}

    \eIf{$\pset{R}$ contains a non-zero constant}{
        $\pset{A}:=\{1\}$; \label{ln:set-a}$\pset{G}:=\{1\}$; {\bf break}\;
    }{
        $\pset{A}:=\pset{A}\setminus\{P,Q\}\cup\pset{R}\setminus\{0\}$\;\label{ln:set-b}
    }

    \If{$P,Q\in\pset{G}$ {\bf and} $b$\label{ln:PQG}}{
        ${\pset{G}}:={\pset{G}\setminus\{P,Q\}\cup\pset{R}}$\;\label{ln:id-rep}
    }\label{ln:end-if}

}

$\pset{M}:=\algbasicset(\pset{A}\cup\pset{F})$\;
\end{algorithm}

\begin{pf}(Algorithm~\ref{alg:autoset})
\emph{Correctness.} For the statement
$[\pset{R},b]:=\algremplus(P,Q,\rd)$ in line \ref{ln:rem}, we have
$\pset{R}\subseteq\bases{P,Q}$ by Definition \ref{def:weakred}. It
is thus easy to verify that $\pset{A}\subseteq\bases{\pset{F}}$
always holds during the running of the algorithm, which means
$\pset{M}\subseteq\bases{\pset{F}}$. Obviously, the ranking of
$\pset{M}$ is not higher than the ranking of any basic set of
$\pset{F}$; hence $\pset{M}$ is a medial set of $\pset{F}$ by
definition.

Let $\pset{G}_i$ be the initial value of {$\pset{G}$} in the $i$th {\bf while} loop.
Obviously, $\bases{\pset{G}_1}=\bases{\pset{F}}$. Suppose that $\bases{\pset{G}_i}=\bases{\pset{F}}$,
and we assert that $\bases{\pset{G}_{i+1}}=\bases{\pset{F}}$ as follows.

Consider the statement $[\pset{R},b]:=\algremplus(P,Q,\rd)$ in line
\ref{ln:rem} of the $i$th {\bf while} loop. If the Boolean
expression ($P,Q\in\pset{G}$ {\bf and} $b$) equals \textbf{false},
then $\pset{G}_{i+1}=\pset{G}_i$. Hence
$\bases{\pset{G}_{i+1}}=\bases{\pset{F}}$. Now assume that
($P,Q\in\pset{G}$ {\bf and} $b$) equals \textbf{true}. Then line
\ref{ln:id-rep} is executed, and we have
$\pset{G}_{i+1}=\pset{G}_i\setminus\{P,Q\}\cup\pset{R}$. As $\rd$ is
an admissible reduction, we have $\pset{R}\subseteq\bases{P,Q}$. By
the assumption on $b$ returned by $\algremplus$, one has
$P,Q\in\bases{\pset{R}}$. Thus $\bases{P,Q}=\bases{\pset{R}}$, which
implies that
$\bases{\pset{G}_{i+1}}=\bases{\pset{G}_i\setminus\{P,Q\}\cup\pset{R}}$.
From the above, $\bases{\pset{G}_{i+1}}=\bases{\pset{F}}$, so
$\bases{\pset{G}}=\bases{\pset{F}}$ always holds.

\emph{Termination.} It is obvious by the assumption on $\textsf{cond}$.
\end{pf}

\begin{pf}(Algorithm~\ref{alg:charset})
\emph{Correctness.} From the properties of $\algautoset$'s output,
we know that $\bases{\pset{B}}=\bases{\pset{G}}$ in line
\ref{ln:autoset}, and $\pset{A}$ is a medial set of $\pset{G}$. It
follows that $\pset{A}\subseteq\bases{\pset{G}}$. According to the
pseudo-division formula, for any polynomial $R$ in $\pset{R}$ of
line \ref{ln:cs-prem}, there exist $B\in\pset{B}\setminus\pset{A}$
and $G_i\in\kx$ such that
$$(\prod_i\ini(A_i)^{s_i})\cdot B=\sum_i G_iA_i +R,$$
where each $s_i$ is a non-negative integer and $A_i\in\pset{A}$.
Hence $\pset{R}\subseteq\bases{\pset{G}}$ always holds during the
running of the algorithm. Thus the ideal generated by $\pset{G}$
remains the same after the assignment in line \ref{ln:cs-g}, which
means that $\bases{\pset{G}}=\bases{\pset{F}}$ always holds.

On the other hand, $\prem(\pset{G},\pset{A})=\{0\}$ when the {\bf while} loop terminates.
Hence $\pset{A}$ is a generalized characteristic set of $\pset{F}$ by Definition \ref{def:gencharset}.

\emph{Termination.} We use $\pset{A}_i$ and $\pset{G}_i$ to denote
the values of $\pset{A}$ and $\pset{G}$ respectively
 in the $i$th {\bf while} loop after executing line \ref{ln:autoset}.
Recalling the properties of $\algautoset$, we know that each
$\pset{A}_i$ is a medial set of $\pset{G}_i$. By Proposition
\ref{prop:med-prec}, one can obtain the sequence
$\pset{A}_1\succ\pset{A}_2\succ\pset{A}_3\succ\cdots$ of triangular
sets, which should be finite by Proposition \ref{prop:zhi-finite}.
Thus the algorithm terminates.
\end{pf}

\section{Concrete Admissible Reductions}

In this section we introduce and discuss several concrete admissible
reductions.

$\rd_{\rm UG}$~(\emph{univariate GCD reduction}): define $\algrem(P,Q,\rd_{\rm UG}):=$
\begin{equation*}
\left\{ \begin{aligned}
         &[0,\gcd(P,Q,x_q)], && \text{if $P,Q$ are univariate polynomials in $x_q$}; \\
         &[P,Q], && \text{otherwise}.
         \end{aligned} \right.
\end{equation*}
In the above definition, $x_q$ is some variable in $\xvar$. It is
easy to verify that $\rd_{\rm UG}$ is an admissible reduction, and
$P$ is $\rd_{\rm UG}$-reducible w.r.t.\ $Q$ if and only if $P, Q$
are univariate polynomials in the same variable.

Pseudo-division, used frequently in triangular decomposition, is
also an admissible reduction.

$\rd_{\rm P}$~(\emph{pseudo-division reduction}):
define~$\algrem(P,Q,\rd_{\rm P}):=[\prem(P,Q,\lv(Q)),Q]$. We have
the following pseudo-division formula
\begin{equation}\label{eq:weichu}
  \ini(Q)^sP=\pquo(P,Q,\lv(Q))\,Q+\prem(P,Q,\lv(Q)),
\end{equation}
where $s$ is a non-negative integer. Thus $\algrem(P,Q,\rd_{\rm
P})\subseteq\bases{P,Q}$, which means that $\rd_{\rm P}$ is an
admissible reduction in $\kx$.

\begin{prop}
  $P$ is $\rd_{\rm P}$-reducible w.r.t.\ $Q$ if and only if $P$ is reducible w.r.t.\ $Q$.
\end{prop}
\begin{pf}
Let $x_q=\lv(Q)$.

\chongfen Suppose that $P$ is reduced w.r.t.\ $Q$. According to the
definition of pseudo-division, we have $\prem(P,Q,x_q)=P$. Thus $P$
is $\rd_{\rm P}$-reduced w.r.t.\ $Q$.

\biyao Suppose that $P$ is reducible w.r.t.\ $Q$. Then either
$$\prem(P,Q,x_q)=0 \quad \text{or} \quad\deg(\prem(P,Q,x_q),x_q)<\ldeg(Q).$$
If $\prem(P,Q,x_q)=0$, then $P$ is obviously $\rd_{\rm P}$-reducible
w.r.t.\ $Q$. Now assume that $\prem(P,Q,x_q)\neq 0$. By analyzing
the algorithm of pseudo-division (see, e.g.,
\citealt[\textsc{PseudoDivisionRec}]{Mis1993A}), one may find that
for any $i~(q<i\leq n)$, $\deg(\prem(P,Q,x_q),x_i)\leq \deg(P,x_i)$.
As $P$ is reducible w.r.t.\ $Q$,
$$\deg(\prem(P,Q,x_q),x_q)<\ldeg(Q)\leq \deg(P,x_q).$$
Hence $\prem(P,Q,x_q)<P$. It follows that $P$ is $\rd_{\rm
P}$-reducible w.r.t.\ $Q$.
\end{pf}

Pseudo-division can be done step by step using the following
operation.

\begin{defn}
Let $P,Q\in\kx\setminus\nf$ with $x_q=\lv(Q)$, $I=\ini(Q)$, and
$J=\lc(P,x_q)$ and suppose that $P$ is reducible w.r.t.\ $Q$. Then
we can perform the following \emph{one-step pseudo-division}:
\begin{equation}\label{eq:sprediv}
 R:=FP-GQ\,x_q^{\deg(P,x_q)-\ldeg(Q)},
\end{equation}
where $F=\lcm(I,J)/J$ and $G=\lcm(I,J)/I$. $R$ is called the \emph{one-step pseudo-remainder} of $P$
w.r.t.\ $Q$ and denoted by $\stprem(P,Q)$.
\end{defn}

One can see that pseudo-division is recursive application of
one-step pseudo-divisions and thus may immediately lead to big
superfluous factors of the pseudo-remainder. In contrast, it is
easier to control the selection and size of polynomials for one-step
pseudo-division, which may result in smaller reduction-rests when
combined with other reductions.

$\rd_{\rm SP}$~(\emph{one-step pseudo-division reduction}): define $\algrem(P,Q,\rd_{\rm SP}):=$
\begin{equation*}
\left\{ \begin{aligned}
         &[\stprem(P,Q),Q], && \text{if $P$ is reducible w.r.t.\ $Q$}; \\
         &[P,Q], && \text{otherwise}.
         \end{aligned} \right.
\end{equation*}
If $P$ is reducible w.r.t.\ $Q$, then  $\stprem(P,Q)\in\bases{P,Q}$
by \eqref{eq:sprediv}, and hence $\algrem(P,Q,\rd_{\rm
SP})\subseteq\bases{P,Q}$; otherwise, $\algrem(P,Q,\rd_{\rm
SP})=[P,Q]\subseteq\bases{P,Q}$ is obvious. Therefore $\rd_{\rm SP}$
is also an admissible reduction by definition.

\begin{prop}
  $P$ is $\rd_{\rm SP}$-reducible w.r.t.\ $Q$ if and only if $P$ is reducible w.r.t.\ $Q$.
\end{prop}
\begin{pf}
\chongfen If $P$ is reduced w.r.t.\ $Q$, then $\algrem(P,Q,\rd_{\rm
SP})=[P,Q]$, and hence $P$ is $\rd_{\rm SP}$-reduced w.r.t.\ $Q$.

\biyao Suppose that $P$ is reducible w.r.t.\ $Q$. Then
$\algrem(P,Q,\rd_{\rm SP})=[\stprem(P,Q),Q]$. If $\stprem(P,Q)=0$,
then $P$ is obviously $\rd_{\rm SP}$-reducible w.r.t.\ $Q$. Now
assume that $\stprem(P,Q)\neq 0$ and consider \eqref{eq:sprediv}. We
know that
$$F=\lcm(I,J)/J=I/\gcd(I,J)\in\nf[x_1,\ldots,x_{q-1}].$$
Thus for any $i~(q<i\leq n)$, $\deg(FP,x_i)=\deg(P,x_i)$.
Furthermore,
$$\deg(GQ\,x_q^{\deg(P,x_q)-\ldeg(Q)},x_i)=\deg(G,x_i)=\deg(J/\gcd(I,J),x_i)\leq\deg(P,x_i).$$
Hence $\deg(\stprem(P,Q),x_i)\leq\deg(P,x_i)$. Moreover, we have
$$\deg(FP,x_q)=\deg(GQ\,x_q^{\deg(P,x_q)-\ldeg(Q)},x_q),$$
which implies that $\deg(\stprem(P,Q),x_q)<\deg(P,x_q)$. It follows
that $\stprem(P,Q)<P$, so $P$ is $\rd_{\rm SP}$-reducible w.r.t.\
$Q$.
\end{pf}

The division operation, which is used in the computation of
\grobner~bases, can also be viewed as an admissible reduction.
Instead of introducing the division reduction directly, we discuss
the one-step division reduction first and then use it to recursively
define the former.

$\rd_{\rm SD}$~(\emph{one-step division reduction}): define $\algrem(P,Q,\rd_{\rm SD}):=$
\begin{equation*}
\left\{ \begin{aligned}
         &[P-M/\lt(Q)\cdot Q,Q], && \text{if there exists a monomial $M$ of $P$ such that $\lt(Q)\mid M$}; \\
         &[P,Q], && \text{otherwise}.
         \end{aligned} \right.
\end{equation*}
It is easy to verify that $\rd_{\rm SD}$ is an admissible reduction,
and $P$ is $\rd_{\rm SD}$-reducible w.r.t.\ $Q$ if and only if there
exists a monomial of $P$ which can be divided by $\lt(Q)$.

Bearing in mind the relation between $\rd_{\rm SP}$ and $\rd_{\rm
P}$, we can define the \emph{division reduction} $\rd_{\rm D}$ as
follows. First, set $R_0:=P$ and $S_0:=Q$. Then recursively compute
$[R_{i+1},S_{i+1}]:=\algrem(R_i,S_i,\rd_{\rm SD})$, until an integer
$m$ is found such that $R_{m+1}=R_m$. Define $\algrem(P,Q,\rd_{\rm
D}):=[R_m,S_m]$, where $S_m=Q$. Similar to $\rd_{\rm SD}$, $\rd_{\rm
D}$ is an admissible reduction, and $P$ is $\rd_{\rm D}$-reducible
w.r.t.\ $Q$ if and only if there exists a monomial of $P$ which can
be divided by $\lt(Q)$.

By computing subresultant PRS, one can also design a useful
admissible reduction. We give the definition here and the reader may
refer to \cite{GL2003S} for more details. Suppose that
$\lv(P)=\lv(Q)$, {$\ldeg(P)\geq\ldeg(Q)$} and treat $P,Q$ as univariate polynomials in
$x_q=\lv(Q)$ with coefficients in $\field{K}[x_1,\ldots,x_{q-1}]$.
Define the \emph{subresultant PRS} of $P$ and $Q$ w.r.t.\ $x_q$ to be
\begin{equation*}\label{}
       P_1:=P,\quad P_2:=Q,\quad P_{i+2} :=\frac{\prem(P_i,P_{i+1},x_q)}{Q_{i+2}},~ 1\leq i\leq r-2,
\end{equation*}
where %$Q_3:=(-1)^{d_1-d_2+1}$, $H_3:=-1$,
\begin{equation*}
    \begin{split}
      d_i &:=\deg(P_i,x_q),~~i=1,\ldots,r,\\
      Q_3 &:=(-1)^{d_1-d_2+1},\quad H_3:=-1,\\
      Q_i &:=-\lc(P_{i-2},x_q)H_i^{d_{i-2}-d_i-1}, ~~i=4,\ldots,r, \\
      H_i &:= (-\lc(P_{i-2},x_q))^{d_{i-3}-d_{i-2}}H_{i-1}^{1-d_{i-3}+d_{i-2}},
      ~~i=4,\ldots,r,
    \end{split}
\end{equation*}
and $\prem(P_{r-1},P_r,x_q)=0$.
The well-known Subresultant Chain Theorem (\citealt[Theorem
7.9.1]{Mis1993A}) indicates the relation between the subresultant
PRS and the Euclidean {PRS} of $P$ and $Q$: $P_i$ is similar to the
element of the same degree in the Euclidean PRS. Furthermore, the
former may have smaller superfluous factors.

We use $\res(P,Q,x_q)$ to denote the \emph{resultant} of $P$ and $Q$ w.r.t.\ $x_q$.
If $\res(P,Q,x_q)=0$, then $P_{r}$ is a greatest common divisor of
$P$ and $Q$ w.r.t.\ $x_q$ (\citealt[Corollary 7.7.9]{Mis1993A}). If
$\res(P,Q,x_q)\neq 0$, then {$\cls(P_{r})<\cls(P)$ and
$\cls(P_{r-1})=\cls(P)$}. Thus $P_{r-1}$ is a greatest common divisor
of $P$ and $Q$ w.r.t.\ $x_q$ under the condition
$P_{r}=0$\footnote{An accurate description requires the use of
evaluation homomorphism, see \citet[Section 7.3.1]{Mis1993A}.};
on the other hand, under the condition $P_{r}\neq 0$, $P_r$ would be a greatest common divisor of $P$ and $Q$.
%\note{$P_r\neq0$ 意味着$P,Q$没有公因子, 这时会出现矛盾列.to jin: P,Q没有公因子是有可能的}
Hence we can introduce the following reduction.

$\rd_{\rm SC}$~(\emph{subresultant PRS reduction}):
define $\algrem(P,Q,\rd_{\rm SC}):=$
\begin{equation*}
\left\{ \begin{aligned}
    &[0,P_{r}],           &&\text{if}~ \lv(P)=\lv(Q),\ldeg(P)\geq\ldeg(Q)\text{ and } \res(P,Q,\lv(Q))=0;\\
    &[P_{r},P_{r-1}], &&\text{if}~ \lv(P)=\lv(Q),\ldeg(P)\geq\ldeg(Q)\text{ and } \res(P,Q,\lv(Q))\neq0;\\
    &[P,Q],                   && \text{otherwise}.
\end{aligned} \right.
\end{equation*}

\begin{example}
Let $P=a\,x^2+b\,x+c$ and $Q=d\,x+e$. Then the subresultant PRS of
$P$ and $Q$ in $x$ is $a\,x^2+b\,x+c, d\,x+e, d^2c-edb+ae^2$.
According to the definition of $\rd_{\rm SC}$, $\algrem(P,Q,\rd_{\rm
SC})={[d^2c-edb+ae^2, d\,x+e]}$.
\end{example}

According to \citet[Lemma 7.7.4]{Mis1993A}, there exist
$A_j,B_j\in\kx$ such that $A_jP+B_jQ=P_j$, i.e.,
$P_j\in\bases{P,Q}$. Thus $\rd_{\rm SC}$ is an admissible reduction.
In addition, it is easy to verify that  $P$ is $\rd_{\rm SC}$-reducible w.r.t.\ $Q$ if and only if $\lv(P)=\lv(Q)$ {and $\ldeg(P)\geq\ldeg(Q)$.}
If $\ldeg(P)<\ldeg(Q)$, then consider $\algrem(Q,P,\rd_{\rm SC})$.

The standard reduction used in Ritt-Wu's algorithm $\algcs$ and its
variants is pseudo-division only, which may lead to superfluous
factors and the swell of polynomial {coefficients}. This is one of
the main causes that {degrades} the performance of Ritt-Wu's algorithm
in many cases. In the {algorithm $\algcharset$} (line
\ref{ln:cs-prem}) other kinds of admissible reductions may also be
used. The condition of admissible reductions is quite {weak}, which
makes our algorithm clearly more flexible than other existing ones
for computing characteristic sets.

The remaining problem is the computation of $b$ in the step
$[\pset{R},b]:=\algremplus(P,Q,\rd)$ of Algorithm \ref{alg:autoset},
where $P$ is $\rd$-reducible w.r.t.\ $Q$.

Let $[R_1,R_2]=\pset{R}$. For $\rd_{\rm UG}$, we have $R_1=0$ and
$R_2=\gcd(P,Q,x_q)$, which implies that
$\bases{P,Q}\subseteq\bases{R_1,R_2}$. For $\rd_{\rm SD}$ and
$\rd_{\rm D}$, $R_2=Q$ and the formula $R_1=P-AQ$ always holds,
where $A$ is a polynomial in $\kx$. Thus it is obvious that
$P\in\bases{R_1,Q}=\bases{R_1,R_2}$ and $Q\in\bases{R_1,R_2}$. Hence
for any of $\rd_{\rm UG}$, $\rd_{\rm SD}$, and $\rd_{\rm D}$, we can
let $\algremplus$ return \textbf{true} as the value of $b$.

For $\rd_{\rm P}$, recall the pseudo-division formula
\eqref{eq:weichu}. Provided that $\ini(Q)^s$ is a non-zero constant,
we also have $P\in\bases{R_1,Q}=\bases{R_1,R_2}$. Thus for the
pseudo-division reduction, we set $b$ to be the value of the Boolean
expression $(\ini(Q)\in\nf~{\bf or}~s=0)$. Similarly, consider
\eqref{eq:sprediv} for $\rd_{\rm SP}$ and set $b$ to be the value of
the Boolean expression $(F\in\nf)$.

Finally for $\rd_{\rm SC}$, set $b=\textbf{false}$ permanently.

\section{A Sample Subalgorithm}\label{sec:samsub}

In this section we provide details about the algorithm
$\algautoset$. The assignment of the Boolean expression
$\textsf{cond}$ is first considered and the termination of
Algorithm~\ref{alg:autoset} guaranteed by $\textsf{cond}$ is proved.
Then the basic idea and a {sample} algorithm for the operation
$\algfind$ are presented. We conclude this section with discussions
about other possible strategies for $\algfind$.

There are several ways to assign $\textsf{cond}$. For example, one
can set $\textsf{cond}{\,=\,}\textbf{false}$. In this case, the
while loop in $\algautoset$ never starts and $\algcharset$ is
identical to Ritt-Wu's algorithm. Another example is to set a
counter {$i$} of the while loop and assign an inequality to
$\textsf{cond}$, say ${i\leq 50}$; then the while loop will run
${50}$ times. The termination of $\algautoset$ in both cases is
obvious. In our implementation, we let $\textsf{cond}$ {be}
$\algfind(\pset{A},\rdset)\neq [\,]$, which means that the while
loop repeats until there is no triple
$[P,Q,\rd]~(P,Q\in\pset{A},\,P\neq Q,\,\rd\in\rdset)$ such that $P$
is $\rd$-reducible w.r.t.\ $Q$. For this case, the termination of
$\algautoset$ is proved as follows.

For any polynomial $R_1$ in $\pset{F}$, if we perform
$[R_2,Q_2]:=\algrem(R_1,Q_1,\rd)$, where $Q_1$ is chosen by
$\algfind$ such that $R_1$ is $\rd$-reducible w.r.t.\ $Q_1$, then
$R_2$ can be viewed as the successor of $R_1$ with $R_1>R_2$; if we
perform $[P_2,R_2]:=\algrem(P_1,R_1,\rd')$, where $P_1$ is chosen by
$\algfind$ such that $P_1$ is $\rd'$-reducible w.r.t.\ $R_1$, then
$R_2$ can be viewed as the successor of $R_1$ with $R_1\gtrapprox
R_2$. As the algorithm runs, the successor $R_3$ of $R_2$ and then
the successor $R_4$ of $R_3$ may be produced in the same way. Thus
one may obtain a polynomial sequence $R_1\gtrapprox R_2 \gtrapprox
R_3 \gtrapprox\cdots$. By Lemma \ref{lem:fseq}, there exists an
integer $m$ such that $R_m\thickapprox R_{m+1} \thickapprox \cdots$.
As this is the case for every polynomial in $\pset{F}$, the operation $\algfind$ will find no triple in the end. Hence the
algorithm terminates.

{Now} we explain the basic idea of the algorithm $\algfind$. The set
of selected admissible reductions is $\rdset=\{\rd_{\rm UG},\rd_{\rm
SD},\rd_{\rm SC},\rd_{\rm SP}\}$\footnote{$\rd_{\rm D}$ and
$\rd_{\rm P}$ are omitted because they can be treated as {recursion}
of $\rd_{\rm SD}$ and $\rd_{\rm SP}$ respectively.}. There may exist
several triples and/or several admissible reductions between a pair
of polynomials. We would like to select better triples first.
Therefore, a way to measure the goodness of triples should be
defined. In addition, an order of the admissible reductions in
$\rdset$ is also needed. We give a sample order on triples as
follows.

First of all, it is natural that a triple with $\rd_{\rm UG}$ is better than other triples without it.
In fact, several univariate pseudo-remainders may be produced in
line 1.7 of the algorithm $\algcharset$. It is easier to deal with
the {$\rd_{\rm UG}$ reduction-rest} of such pseudo-remainders in the
next while loop than to deal with them immediately. Next, a triple
with $\rd_{\rm SD}$ should be better than a triple with $\rd_{\rm
SP}$ or $\rd_{\rm SC}$. There are two reasons for this.
\begin{enumerate}
  \item[(1)] For two triples $[P,Q,\rd_{\rm SD}]$ and $[P,Q,\rd_{\rm SP}]$, $P$ is both $\rd_{\rm SD}$-reducible and $\rd_{\rm SP}$-reducible w.r.t.\ $Q$. It is easy to see that computing the $\rd_{\rm SD}$ reduction-rest $R$ of $P$ by $Q$ requires less multiplications than computing the $\rd_{\rm SP}$ reduction-rest $R'$ and in general $R$ is of smaller size than $R'$. Therefore, $[P,Q,\rd_{\rm SD}]$ is better than $[P,Q,\rd_{\rm SP}]$.
  \item[(2)] For two triples $[P,Q,\rd_{\rm SD}]$ and $[P',Q',\rd_{\rm SP}]$ with $P\neq P'$ or $Q\neq Q'$, the former is better because $P$ is in $\bases{Q,R}$, while in general $P'$ is not in $\bases{Q',R'}$, where $R$ and $R'$ are the $\rd_{\rm SD}$ reduction-rest of $P$ by $Q$ and the $\rd_{\rm SP}$ reduction-rest of $P'$ by $Q'$ respectively.
\end{enumerate}
Then, $[P,Q,\rd_{\rm SC}]$ may be better than $[P,Q,\rd_{\rm SP}]$
because the reduction-rest of the former generally involves
coefficients of smaller size than that of the latter. Finally, a
triple $[P,Q,\rd]$ might be better than $[P',Q',\rd]~(\rd\in\rdset)$
if $P>P'$ or $Q<Q'$. This observation is based on the way of
reduction in Ritt-Wu's algorithm: the polynomial of maximal class and maximal degree is first reduced by a polynomial of minimal degree. To summarize, we provide the formal definition of an order on the triples: $[P,Q,\rd]<[P',Q',\rd']$ if $\rd<\rd'$,
or $\rd=\rd'$ and $P<P'$, or $\rd=\rd'$ and $P\thickapprox P'$ and $Q>Q'$. The admissible reductions in $\rdset$ are ordered as $\rd_{\rm UG}>\rd_{\rm SD}>\rd_{\rm SC}>\rd_{\rm SP}$.

Taking as input a polynomial set $\pset{F}$ and a set $\rdset$ of
selected admissible reductions under the above order, $\algfind$
will do the following steps to select a better triple or output
$\emptyset$ if there is none.
\begin{enumerate}
  \item[1.]\label{step:a} Check if there exist triples $[P,Q,\rd_{\rm UG}]~(P,Q\in\pset{F},\,P\neq Q)$. If there is only one, then return it; if there are many, then return the one in which $P$ has maximal class and maximal leading degree and $Q$ has fewest terms and minimal leading degree.
  \item[2.]\label{step:b} Sort the polynomials in $\pset{F}$ increasingly w.r.t.\ the partial order $<$.
  Start with a polynomial $P$ of highest order and check if there exists any polynomial
  $Q$ such that $P$ is $\rd_{\rm SD}$-reducible w.r.t.\ $Q$. If there is only one $Q$,
  then return $[P,Q,\rd_{\rm SD}]$; if there are many, then select $Q$ from
  $$\{F\in\pset{F}:\,P\text{ is }\rd_{\rm SD}\text{-reducible w.r.t.\ }F\}$$
  which has fewest terms and minimal leading degree and return $[P,Q,\rd_{\rm SD}]$;
  otherwise, consider another $P$ of lower order and check again.
  If there is no such triple, then go to the next step. % with $\rd_{\rm SD}$
  \item[3.] Start with a polynomial $P$ of highest order again
  and check if there exists any $Q$ such that $P$ is $\rd_{\rm SC}$-reducible w.r.t.\ $Q$. The process is similar to step \ref{step:b}.
       \item[4.] Start with a polynomial $P$ of highest order again and check if there exists
   any $Q$ such that $P$ is $\rd_{\rm SP}$-reducible w.r.t.\ $Q$. The process is similar to step \ref{step:b}.
   \item[5.] If there is no triple at all, then output $\emptyset$.
\end{enumerate}

The above process is described formally as Algorithm \ref{alg:find}.
Its correctness follows from the above analysis and its termination
is obvious. If ``$P_i \text{ is } \rd_{\rm SP} \text{-reducible w.r.t. } Q$''  in line 3.24 is replaced by ``$\ini(P_i) \text{ is } \rd_{\rm SP} \text{-reducible w.r.t. } Q$'',
then $\algautoset$ computes a weak-ascending set and therefore $\algcharset$ outputs a weak-characteristic set.

\begin{algorithm}[]\label{alg:find} %\caption{$[P,Q,\rd]:=\algfind(\pset{F},\rdset)$}
\KwIn{$\pset{F}$ --- set of polynomials in $\kx$; \linebreak
$\rdset=[\rd_{\rm UG},\rd_{\rm SD},\rd_{\rm SC},\rd_{\rm SP}]$.}

\KwOut{ $\emptyset$ or $[{P},Q,\rd]$ such that $P$ is $\rd$-reducible w.r.t.\ $Q$.%;\linebreak
}
    \BlankLine
    $\pset{S}:=\emptyset$;

    {\bf for} $i=n$ \emph{\KwTo}$1$\While{$|\pset{S}|<2$}
    {$\pset{S}:=\{P\in\pset{F}:\,P\in\field{K}[x_i]$\};}

    \If{$\exists~ P,Q\in\pset{S}$, $P\neq Q$, such that $P$ is $\rd_{\rm UG}$-reducible w.r.t.\ $Q$}{\Return $[P,Q,\rd_{\rm UG}]$ such that $P\in\pset{S}$ has maximal degree, $Q\in\pset{S}\setminus\{P\}$ has fewest terms and minimal degree, and $P$ is $\rd_{\rm UG}$-reducible w.r.t.\ $Q$;}

    $[P_1,\ldots,P_r]:=\pset{F}$\, (with $P_i$ sorted increasingly w.r.t.\ $<$)\; %$\pset{G}:=\{P\in\pset{F}\}$\;
    \For{$i=r$ \emph{\KwTo}$2$}
    {   $\pset{Q}:=\{Q\in\pset{F}\setminus\{P_i\}:\,P_i \text{ is } \rd_{\rm SD} \text{-reducible w.r.t. } Q \}$\;
        \If{$\pset{Q}\neq\emptyset$}
            {choose $Q\in\pset{Q}$ with fewest terms and minimal leading degree\;
             \Return $[P_i,Q,\rd_{\rm SD}]$;}
    }
        \For{$i=r$ \emph{\KwTo}$2$}
    {   $\pset{Q}:=\{Q\in\pset{F}\setminus\{P_i\}:\,P_i \text{ is } \rd_{\rm SC} \text{-reducible w.r.t. } Q \}$\;
        \If{$\pset{Q}\neq\emptyset$}
            {choose $Q\in\pset{Q}$ with fewest terms and minimal leading degree\;
             \Return $[P_i,Q,\rd_{\rm SC}]$;}
    }
        \For{$i=r$ \emph{\KwTo}$2$}
    {   $\pset{Q}:=\{Q\in\pset{F}\setminus\{P_i\}:\,P_i \text{ is } \rd_{\rm SP} \text{-reducible w.r.t. } Q \}$\;
        \If{$\pset{Q}\neq\emptyset$}
            {choose $Q\in\pset{Q}$ with fewest terms and minimal leading degree\;
             \Return $[P_i,Q,\rd_{\rm SP}]$;}
    }
    $\emptyset$\; %   }
\end{algorithm}
%\note{Wang: \textcolor{red}{leading} degree in the algorithm ? double-check lines 3.1--3.8  jin: in this case, P,Q are univariate pols, so it does not matter}

Note that the design of $\algfind$ is flexible and can be made more
technical and comprehensive. There are several ways to improve
$\algfind$. One may introduce other admissible reductions by
computing B\'ezout resultants or \gb bases, or using other
techniques (e.g., from linear algebra). If the input set of
$\algfind$ contains polynomials of special form or structure, then
new reduction strategies may be adopted. For example, if there is
one univariate polynomial of low degree or few (say one or two)
terms, then one should use this polynomial first to reduce (and
simplify) other polynomials. The ordering {used in line
3.8} is crucial for the efficiency of $\algautoset$. One may sort
polynomials w.r.t.\ different orderings depending on the admissible
reductions.

\section{Example and Preliminary Experiments}

In this section we present first an illustrative example to compare
the outputs of different algorithms and to show how the algorithm
$\algcharset$ works and then our experimental results on the
performance of $\algcharset$. In what follows, an index tuple
$${\big[}[\deg(P,x_1),\ldots,\deg(P,x_n)],\nops(P), \lm(P),m {\big]}$$
is used to characterize a polynomial $P\in \kx\setminus \field{K}$,
where $\nops(P)$ denotes the number of terms of (expanded) $P$,
$\lm(P)$ the heading monomial of $P$, and
$m$ the maximal number of digits of the integer coefficients of $P$.%\note{m digits?这写错了吧}
\begin{example}[{\citealt[Epsilon-A14]{w04e}}]\label{example:A14}
Let $\pset{F}=\{F_1,F_2,F_3\}\subseteq\qnum[w,x,y,z]$, where%\note{符号规则保持和多项式代数一样}
$$F_1 = x^2+y^2+z^2-w^2,\quad F_2 = xy+z^2-1, \quad F_3 = xyz-x^2-y^2-z+1,$$
%\begin{equation*}\label{}
%    \begin{split}
%    F_1 &= x^2+y^2+z^2-w^2, \\
%    F_2 &= xy+z^2-1,   \\
%    F_3 &= xyz-x^2-y^2-z+1,
%     \end{split}
%\end{equation*}
with $w<x<y<z$. The output of $\algcs$\footnote{Corresponding to the Maple command charset($\pset{F},[w,x,y,z]$) with the Epsilon package (\citealt{w04e}).} consists of three polynomials with index tuples
\begin{equation}\label{eq:example1output1}
    \begin{split}
    &\big[[520, 42, 0, 0], 5450,  w^{508}x^{42}, 212\big], \\
    &\big[[6, 30, 1, 0], 93, {x^{29}y}, 3 \big], \\
    &\big[[520, 40, 0, 1], 10390, w^{510}x^{40}z, 212\big],
     \end{split}
\end{equation}
while the output of $\algcharset$ also consists of three polynomials
yet with index tuples
\begin{equation}\label{eq:example1output}
  \big[[8, 12, 0, 0], 23,  x^{12}, 2\big], \quad \big[[4, 6, 1, 0], 12, {w^2x^3y}, 1 \big], \quad \big[[4, 6, 0, 1], 17, x^6z, 1\big].
\end{equation}
%\begin{equation*}\label{}
%    \begin{split}
%    &[[8, 12, 0, 0], 23,  x^{12}, 2], \\
%    &[[4, 6, 1, 0], 12, yx^3w^2, 1 ],\\
%    &[[4, 6, 0, 1], 17, x^6z, 1].
%     \end{split}
%\end{equation*}
%\note{用数字, 全文检查 jin:按照on the theory of tri sets 的写法是用单词 Wang: 大个数用数字}
The \gb basis of $\pset{F}$ w.r.t.\ the lexicographic order
(determined by $w<x<y<z$) contains $9$ polynomials, of which one of the biggest has index tuple ${\big[[12, 11,
1, 1],41,yz,5\big]}$.

Now let us show how $\algcharset$ works. For brevity, the admissible
reductions $\rd_{\rm SD}$ and $\rd_{\rm SP}$ are renamed $\rd_{\rm
D}$ and $\rd_{\rm P}$ respectively. After the initialization of
$\pset{G}$ and $\pset{A}$ in line~1.1 of $\algcharset$, the first
{\bf while} loop (line~1.2) starts and
$\algautoset(\pset{F},\rdset)$ is called. Line~2.1 initializes the
values of $\pset{A}$ and $\pset{G}$. Then it is checked whether the
Boolean expression $\algfind(\pset{A},\rdset)\neq [\,]$ is
\textbf{true} or \textbf{false}. One sees clearly that $F_3<F_2<F_1$
and there are three triples
$$[F_2,F_3,\rd_{\rm D}],\quad [F_1,F_3,\rd_{\rm D}],\quad [F_1,F_2,\rd_{\rm D}].$$
Therefore the first {\bf while} loop (line 2.2) starts.
$\algfind(\pset{A},\rdset)$ outputs the third triple and
$\algremplus(F_1,F_2,\rd_{\rm D})$ then returns
$[\algrem(F_1,F_2,\rd_{\rm D}), \textbf{true}]$. Since the
reduction-rest contains no constant, $\pset{A}$ and $\pset{G}$ are
updated and the second {\bf while} loop (line 2.2) starts. After the
second {\bf while} loop, $\pset{G}$ is updated to be
\begin{equation}\label{eq:example1output2}
  \{y^2-{xy}+x^2-w^2+1, {xyz-z-xy}-w^2+2,z^2+{xy}-1\}
\end{equation}%\note{wang 超过3的序数,用数字?}
and remains unchanged afterwards. After the 7th {\bf while} loop,
$\pset{A}$ is updated to be an ascending set with index tuples of
its polynomials shown in \eqref{eq:example1output} and
$\algfind(\pset{A},\rdset)=[\,]$. So $\algautoset$ returns a basic
set of $\pset{A}\cup\pset{F}$ (which is $\pset{A}$) as well as
\eqref{eq:example1output2}. It can be verified by line 1.7 that the
pseudo-remainders of the polynomials in \eqref{eq:example1output2}
w.r.t.\ $\pset{A}$ are all zero. Note that it is easier to compute
$\prem(\pset{G}\setminus\pset{A},\pset{A})$ than
$\prem(\pset{F}\setminus\pset{A},\pset{A})$ because one polynomial
in \eqref{eq:example1output2} is of smaller class than the
polynomials in $\pset{F}$. Therefore, $\pset{A}$ with the index
tuples in \eqref{eq:example1output} for its three polynomials is a
characteristic set of $\pset{F}$.
\end{example}

We have implemented the algorithm $\algcharset$ described in Section
3 using the Epsilon library. There are two versions of it:
$\algcharsetw$ outputs weak-characteristic sets and $\algcharset$
outputs characteristic sets. We have compared the performances of
$\algcharset$, $\algcharsetw$, $\algcs$, and $\algcsw$\footnote{By
$\algcs$ and $\algcsw$ we mean the algorithms implemented in Epsilon
which adopted optional strategies from \cite{Wang1992A,Wang2001A}
for speeding up the computation of characteristic sets rather than
Ritt-Wu's original algorithm.}as well as the Gr\"{o}bner basis
algorithm $\alggb$\footnote{Corresponding to the Maple command
Groebner[Basis]($\pset{F}$, plex($x_n,\ldots,x_1$)).}. The timings
(in CPU seconds) given in Table \ref{tab:timing} are for our
implementation running in Maple {14} under Windows 7 on
a laptop Intel Core 2 Duo T6670 Processor 2.20\,GHz with 3\,GB of
memory. The asterisk * indicates {that the
computation is} out of memory. Table \ref{tab:nops} shows { the size of the test examples (including the number of variables and the total degree of each polynomial) and} the number of polynomials in the
output of each algorithm. {The entry $[3, 3, 4]$ in the table means that example 1 consists of three polynomials whose total degree are $3,3,4$ respectively.} The test examples are taken from
http://www.symbolicdata.org/ and
http://www-salsa.lip6.fr/\textasciitilde wang/epsilon/. It is easy
to see that $\algcharset$ is more efficient than $\algcs$ and
$\algcsw$ for most of the examples and is comparable with $\alggb$.
\begin{table}[]\footnotesize
\centering
\caption{Timings (in seconds) for five algorithms}
\begin{tabular}{clccccc}
\toprule
    No&   Name        &$\algcsw$  & $\algcs$    &$\algcharsetw$&$\algcharset$& $\alggb$    \\
\midrule    %\hline
    1 & DiscrC2       & 0.016     & 0.016       & 0.016        & 0.016       &  $>$1000 \\\hline
    2 & Epsilon-A14   & $>$1000   & 16.302      & 0.016        & 0.031       & 0.156  \\\hline
    3 & Chou156-1     & 0.218     &  0.093      & 0.047        &  0.156      &19.812   \\\hline
    4 & Trinks1       &$>$1000    &$>$1000      &0.063         &  0.063      & 0.047   \\\hline
    5 & ZeroDim14     & 34.398    & 0.795       &  0.156       & 7.909       & 0.172  \\\hline
    6 & Epsilon-A16   &  761.145  &  8.128      &  0.390       & 3.073       & 0.266  \\\hline
    7 & Schiele1      & 1.576     & 1.482       & 0.406        & 0.375       & 5.491  \\\hline
    8 & Epsilon-A25   & 0.172     &  0.078      &  0.546       &  7.301      & 1.029  \\\hline
    9 & Cyclic5       & $>$1000   & $>$1000     &  0.577       & $>$1000     &  0.062 \\\hline
    10& Fee1          & 676.061   & 2.777       &  1.326       & 5.273       & 0.156  \\\hline
    11& Weispfenning94& $>$1000   & 33.680      & 2.340        & 7.238       & 0.921  \\\hline
    12& Steidel2      & $>$1000   & 49.312      & 5.475        &15.678       & 0.905  \\\hline
    13& Epsilon-A30   & $>$1000    & $>$1000      & 7.941        & $>$1000      & 2.450  \\\hline % 3 >500
    14& Fateman       & $>$1000   & $>$1000     & 9.797        &24.429       &411.546\\\hline
    15& Sym3-5        & 23.868    & 1.623       & 10.733       &10.483       &  35.865\\\hline
    16& Epsilon-A3    & 160.150   & 67.174      & 17.035       & 261.192     & *    \\\hline
    %17% Rose         & 656.437   & 21.840      & 11.092       &  7.519      & 7.566  \\
    17& Wu90          & $>$1000   & $>$1000     & 44.835       &  $>$1000    & *  \\\hline
    18& Cyclic6       & $>$1000   & $>$1000     & $>$1000      & $>$1000     & 0.374\\
\bottomrule
\end{tabular}\label{tab:timing}
 \end{table}
\begin{table}[]\footnotesize
\centering \caption{Numbers of polynomials in the outputs of five
algorithms}
\begin{tabular}{ccccccccc}
\toprule
    No & {No of vars}& {Total degree list} &$\algcsw$  & $\algcs$    &$\algcharsetw$&$\algcharset$& $\alggb$    \\
\midrule
    1  & 12 & [3, 3, 4]       & 3   & 3     & 3    & 3   &   \\\hline      % & DiscrC2
    2  & 4  & [1, 2, 3, 4]     &    & 3    & 3    & 3   & 9  \\\hline       % & Epsilon-A14
    3  & 4  & [2, 2, 2, 2]  & 4   &  4    & 4    &  4  &49   \\\hline    % & Chou156-1
    4  & 6  & [1, 1, 2, 2, 2, 3]    &    &      &6     &  6  & 6   \\\hline      % & Trinks1
    5  & 4  & [1, 2, 3, 4]    & 4  & 4     &  4   & 4   & 4  \\\hline      % & ZeroDim14
    6  & 4  & [4, 3, 2, 4]   &  4  &  4    &  4   & 4   & 4  \\\hline     % & Epsilon-A16
    7  & 2  & [6, 11]    & 2   & 2     & 2    & 2   & 3  \\\hline     % & Schiele1
    8  & 8  & [2, 2, 2, 2]    & 4   &  4    &  4   &  4  & 21  \\\hline    % & Epsilon-A25
    9  & 5  & [1, 2, 3, 4, 5]    &    &      &  5   &   &  11 \\\hline        % & Cyclic5
    10 & 4  & [2, 3, 4, 4]    & 4 & 4     &  4   & 4   & 5  \\\hline       % & Fee1
    11 & 3  & [4, 5, 5]    &    & 3    & 3    & 3   & 3  \\\hline       % & Weispfenning94
    12 & 3  & [4, 5, 5]   &    & 3    & 3    &3   & 3  \\\hline        % & Steidel2
    13 & 10  & [3, 3, 3, 3, 3, 3, 3, 3, 3, 3]    &     &       &10   &     & 10  \\\hline     % & Epsilon-A30
    14 & 4 &  [3, 5, 5]   &    &      & 3    &3   &108\\\hline         % & Fateman
    15 & 3  & [6, 6, 6]    & 3  & 3     & 3   &3   &  3\\\hline         % & Sym3-5
    16 & 4   & [4, 4, 4, 5]     & 4 & 4    & 4   & 4 &     \\\hline          % & Epsilon-A3
    17 & 4  &  [3, 3, 3, 4]   &    &      & 4   &    &   \\\hline          % & Wu90
    18 & 6   &  [1, 2, 3, 4, 5, 6]   &    &      &     &    & 17\\                % & Cyclic6
\bottomrule
\end{tabular}\label{tab:nops}
 \end{table}

Now let us comment on the empirical results shown in Table
\ref{tab:timing}. For examples $3$, $5$, $8$, $10$, $15$, and $16$,
$\algcharset$ is slower than $\algcs$, and for examples $9$, $13$,
$17$, and $18$, neither of them can get any result within 1000
seconds. For the $8$ other examples (notably, examples $4$ and
$14$), $\algcharset$ is always faster than $\algcs$. Among the first
four algorithms, $\algcharsetw$ is the most efficient (for all but
one test example, i.e., example $15$). There are five examples for
which $\algcs$ cannot get any result within 1000 seconds. The
$\alggb$ algorithm is slower than other algorithms for examples $1$,
$3$, $7$, $14$, and $15$, but considerably faster for examples $9$
and $18$. There are two examples for which $\alggb$ cannot complete
the computation due to the lack of memory.

There are several reasons for $\algcharset$ to be more efficient
than $\algcs$. First of all, it is the polynomials in the output
$\pset{G}$ of $\algautoset$ (instead of $\pset{F}$) whose
pseudo-remainders w.r.t.\ the output (weak-) medial set $\pset{M}$
of $\algautoset$ are computed. This may reduce the cost of zero
pseudo-remainder verification in most cases because the polynomials
in $\pset{G}$ are ``closer'' (than those in $\pset{F}$) to the
polynomials in $\pset{M}$. For example, if $\pset{G}$ in
$\algautoset$ is fixed to be $\pset{F}$, then $\algcharset$ takes
42.307 seconds (vs 15.678 seconds when $\pset{G}$ remains updated)
for example 12 and gets no result within 500 seconds (vs 24.429
seconds when $\pset{G}$ remains updated) for example 14. However,
keeping $\pset{G}$ updated may also lower the efficiency of the
algorithm for some examples {(including three of our test examples)}.
How to choose $\pset{G}$ to improve the
efficiency is a question that remains for further investigation.

The use of admissible reductions allows us to effectively control
the swell of polynomial coefficients as well as degrees. As we have
seen, the maximal numbers of digits of coefficients in
\eqref{eq:example1output1} are much bigger
than those in \eqref{eq:example1output}. More comparisons are given
in Table \ref{tab:digits} in the appendix. The ordering $<$ {used in line 3.8
can also be replaced by other orderings}. According to our experiments, there
may be fewer while loops in $\algcharset$ if polynomials are sorted w.r.t.\
the lexicographic order of their degree tuples, {i.e., $P$ is lower than $Q$} if %\m{$P<_{\rm lex}Q$}.
$$[\deg(P,x_1),\ldots,\deg(P,x_n)]<_{\rm lex}[\deg(Q,x_1),\ldots,\deg(Q,x_n)].$$
For example, let $P=yz+x^2z+1$ and $Q=xz+z+y^2$ with $x<y<z$; then $P>Q$ according to
Definition \ref{def:rankofpol}. However, the above ordering indicates that $P$ is lower
than $Q$. One can see that the degree tuples of $P$ and $Q$ are $[2,1,1]$
and $[1,2,1]$ respectively, so $[2,1,1]<_{\rm lex}[1,2,1]$. Note that $x<y<z$ is used instead of $x>y>z$, refer to \cite[Chapter 2, Definition 3 (Lexicographic Order)]{c97i} for the difference.

In the case where the polynomials
in the input set have many (say more than 6) variables, $\alggb$
needs to deal with many intermediate polynomials (see examples
$3,14,16$, and $17$ in Table \ref{tab:nops}) and $\algcs$ and
$\algcsw$ need to deal with polynomials of complex initials and high
degrees, while $\algcharsetw$ can partially avoid such problems.

Finally, we point out that admissible reductions may also be
incorporated into other algorithms of triangular decomposition,
which would create plenty of room for improvements on all such
algorithms.

\bibliographystyle{elsart-harv}
\bibliography{CharSet}

\begin{thebibliography}{42}
\expandafter\ifx\csname natexlab\endcsname\relax\def\natexlab#1{#1}\fi
\expandafter\ifx\csname url\endcsname\relax
  \def\url#1{\texttt{#1}}\fi
\expandafter\ifx\csname urlprefix\endcsname\relax\def\urlprefix{URL }\fi

\bibitem[{Aubry et~al.(1999)Aubry, Lazard, and Moreno~Maza}]{a99t}
Aubry, P., Lazard, D., Moreno~Maza, M., 1999. On the theories of triangular
  sets. Journal of Symbolic Computation 28~(1--2), 105--124.

\bibitem[{Aubry and Moreno~Maza(1999)}]{a99ta}
Aubry, P., Moreno~Maza, M., 1999. Triangular sets for solving polynomial
  systems: a comparative implementation of four methods. Journal of Symbolic
  Computation 28~(1--2), 125--154.

\bibitem[{Boulier et~al.(1995)Boulier, Lazard, Ollivier, and
  Petitot}]{BLOP1995R}
Boulier, F., Lazard, D., Ollivier, F., Petitot, M., 1995. Representation for
  the radical of a finitely generated differential ideal. In: Levelt, A. H.~M.
  (Ed.), Proceedings of the 1995 International Symposium on Symbolic and
  Algebraic Computation. Association for Computing Machinery, New York, pp.
  158--166.

\bibitem[{Boulier et~al.(2009)Boulier, Lazard, Ollivier, and
  Petitot}]{BLFP2009C}
Boulier, F., Lazard, D., Ollivier, F., Petitot, M., 2009. Computing
  representations for radicals of finitely generated differential ideals.
  Applicable Algebra in Engineering, Communication and Computing 20~(1),
  73--121.

\bibitem[{Boulier et~al.(2010)Boulier, Lemaire, and Moreno~Maza}]{BLM2010C}
Boulier, F., Lemaire, F., Moreno~Maza, M., 2010. Computing differential
  characteristic sets by change of ordering. Journal of Symbolic Computation
  45~(1), 124--149.

\bibitem[{Chen and Moreno~Maza(2011)}]{CM2011A}
Chen, C., Moreno~Maza, M., 2011. Algorithms for computing triangular
  decompositions of polynomial systems. In: Leykin, A. (Ed.), Proceedings of
  the 36th International Symposium on Symbolic and Algebraic Computation.
  Association for Computing Machinery, New York, pp. 83--90.

\bibitem[{Chen and Gao(2003)}]{CG2003I}
Chen, Y., Gao, X.-S., 2003. Involutive characteristic sets of algebraic partial
  differential equation systems. Science in China Series A: Mathematics 46~(4),
  469--487.

\bibitem[{Chou(1988)}]{Chou1988M}
Chou, S.-C., 1988. Mechanical Geometry Theorem Proving. Reidel, Dordrecht.

\bibitem[{Chou and Gao(1990)}]{CG1990R}
Chou, S.-C., Gao, X.-S., 1990. {Ritt-Wu}'s decomposition algorithm and geometry
  theorem proving. In: Stickel, M.~E. (Ed.), Proceedings of the 10th
  International Conference on Automated Deduction. Vol. 449 of Lecture Notes in
  Computer Science. Springer-Verlag, Berlin, pp. 207--220.

\bibitem[{Collins(1967)}]{Col1967S}
Collins, G.~E., 1967. Subresultants and reduced polynomial remainder sequences.
  Journal of the Association for Computing Machinery 14~(1), 128--142.

\bibitem[{Cox et~al.(1997)Cox, Little, and O'Shea}]{c97i}
Cox, D.~A., Little, J.~B., O'Shea, D., 1997. {Ideals, Varieties, and
  Algorithms: An Introduction to Computational Algebraic Geometry and
  Commutative Algebra}, 2nd Edition. Undergraduate Texts in Mathematics.
  Springer, New York.

\bibitem[{Gallo and Mishra(1991{\natexlab{a}})}]{GM1991E}
Gallo, G., Mishra, B., 1991{\natexlab{a}}. Efficient algorithms and bounds for
  {Wu-Ritt} characteristic sets. In: Mora, T., Traverso, C. (Eds.), Effective
  Methods in Algebraic Geometry. Vol.~94 of Progress in Mathematics.
  Birkhauser, Berlin, pp. 119--142.

\bibitem[{Gallo and Mishra(1991{\natexlab{b}})}]{GM1991W}
Gallo, G., Mishra, B., 1991{\natexlab{b}}. {Wu-Ritt characteristic sets and
  their complexity}. In: Goodman, J.~E., Pollack, R.~D., Steiger, W. (Eds.),
  Discrete and Computational Geometry: Papers from the DIMACS Special Year.
  Vol.~6 of Discrete Mathematics and Theoretical Computer Science. American
  Mathematical Society, Providence, pp. 111--136.

\bibitem[{Gao and Chou(1992)}]{g92s}
Gao, X.-S., Chou, S.-C., 1992. Solving parametric algebraic systems. In: Wang,
  P.~S. (Ed.), Proceedings of the 1992 International Symposium on Symbolic and
  Algebraic Computation. Association for Computing Machinery, New York, pp.
  335--341.

\bibitem[{Gao and Chou(1993)}]{g93d}
Gao, X.-S., Chou, S.-C., 1993. On the dimension of an arbitrary ascending
  chain. Chinese Science Bulletin 38~(5), 396--399.

\bibitem[{Gao et~al.(2009)Gao, Luo, and Yuan}]{GLC2009A}
Gao, X.-S., Luo, Y., Yuan, C., 2009. A characteristic set method for ordinary
  difference polynomial systems. Journal of Symbolic Computation 44~(3),
  242--260.

\bibitem[{Hubert(2000)}]{Hub2000F}
Hubert, E., 2000. Factorization-free decomposition algorithms in differential
  algebra. Journal of Symbolic Computation 29~(4--5), 641--662.

\bibitem[{Hubert(2003{\natexlab{a}})}]{Hub2003}
Hubert, E., 2003{\natexlab{a}}. Notes on triangular sets and
  triangulation-decomposition algorithms {I}: polynomial systems. In: Winkler,
  F., Langer, U. (Eds.), Symbolic and Numerical Scientific Computation. Vol.
  2630 of Lecture Notes in Computer Science. Springer-Verlag, Berlin, pp.
  143--158.

\bibitem[{Hubert(2003{\natexlab{b}})}]{Hub2003N}
Hubert, E., 2003{\natexlab{b}}. Notes on triangular sets and
  triangulation-decomposition algorithms {II}: differential systems. In:
  Symbolic and Numerical Scientific Computation. Vol. 2630 of Lecture Notes in
  Computer Science. Springer-Verlag, Berlin, pp. 40--87.

\bibitem[{Kalkbrener(1993)}]{k93g}
Kalkbrener, M., 1993. A generalized {Euclidean} algorithm for computing
  triangular representations of algebraic varieties. Journal of Symbolic
  Computation 15~(2), 143--167.

\bibitem[{Kapur and Lakshman(1992)}]{KL1992E}
Kapur, D., Lakshman, Y., 1992. Elimination methods: an introduction. In:
  Donald, B.~R., Kapur, D., Mundy, J.~L. (Eds.), Symbolic and Numerical
  Computation for Artificial Intelligence. Academic Press, San Diego, pp.
  45--89.

\bibitem[{Knuth(1997)}]{Knuth1997T}
Knuth, D.~E., 1997. The Art of Computer Programming, Volume 2 (3rd edn.):
  Seminumerical Algorithms. Addison-Wesley Longman Publishing Company, Boston.

\bibitem[{Lazard(1991)}]{l91n}
Lazard, D., 1991. A new method for solving algebraic systems of positive
  dimension. Discrete Applied Mathematics 33~(1--3), 147--160.

\bibitem[{Li(2006)}]{Li2006S}
Li, Y., 2006. Some properties of triangular sets and improvement upon algorithm
  {CharSer}. In: Calmet, J., Ida, T., Wang, D. (Eds.), Artificial Intelligence
  and Symbolic Computation. Vol. 4120 of Lecture Notes in Computer Science.
  Springer-Verlag, Beijing, pp. 82--93.

\bibitem[{Li(1989)}]{Li1989D}
Li, Z., 1989. Determinant polynomial sequences. Chinese Science Bulletin 34,
  1595--1599.

\bibitem[{Li and Wang(1999)}]{LW1999}
Li, Z., Wang, D., 1999. Cohernet, regular and simple systems in zero
  decompositions of partial. Systems Science and Mathematical Sciences 12,
  43--60.

\bibitem[{Mishra(1993)}]{Mis1993A}
Mishra, B., 1993. {Algorithmic Algebra}. Texts and Monographs in Computer
  Science. Springer-Verlag, New York.

\bibitem[{Moreno~Maza(2000)}]{m00t}
Moreno~Maza, M., 2000. On triangular decompositions of algebraic varieties.
  {Technical Report 4/99, NAG, UK}, {Presented at the MEGA-2000 Conference,
  Bath}.

\bibitem[{Ritt(1950)}]{r50d}
Ritt, J.~F., 1950. {Differential Algebra}. Vol.~33 of Colloquium Publications.
  American Mathematical Society, New York.

\bibitem[{Sasaki and Furukawa(1984)}]{SAS1984T}
Sasaki, T., Furukawa, A., 1984. Theory of multiple polynomial remainder
  sequence. Publications of the Research Institute for Mathematical Sciences
  20~(2), 367--399.

\bibitem[{Szanto(1999)}]{Sza1999C}
Szanto, A., 1999. Computation with polynomial systems. Ph.D. thesis, Cornell
  University, Ithaca.

\bibitem[{von~zur Gathen and Lucking(2003)}]{GL2003S}
von~zur Gathen, J., Lucking, T., 2003. Subresultants revisited. Theoretical
  Computer Science 297~(1--3), 199--239.

\bibitem[{Wang(1992)}]{Wang1992A}
Wang, D., 1992. A strategy for speeding-up the computation of characteristic
  sets. In: Havel, I.~M., Koubek, V. (Eds.), Proceedings of the 17th
  International Symposium on Mathematical Foundations of Computer Science. Vol.
  629 of Lecture Notes in Computer Science. Springer-Verlag, Berlin, pp.
  504--510.

\bibitem[{Wang(1993)}]{w93e}
Wang, D., 1993. An elimination method for polynomial systems. Journal of
  Symbolic Computation 16~(2), 83--114.

\bibitem[{Wang(1995)}]{Wang1995A}
Wang, D., 1995. An implementation of the characteristic set method in {Maple}.
  In: Pfalzgraf, J., Wang, D. (Eds.), Automated Practical Reasoning: Algebraic
  Approaches. Texts and Monographs in Symbolic Computation. Springer-Verlag,
  Wien New York, pp. 187--201.

\bibitem[{Wang(2000)}]{w00c}
Wang, D., 2000. Computing triangular systems and regular systems. Journal of
  Symbolic Computation 30~(2), 221--236.

\bibitem[{Wang(2001{\natexlab{a}})}]{Wang2001A}
Wang, D., 2001{\natexlab{a}}. {A generalized algorithm for computing
  characteristic sets}. In: Shirayanagi, K., Yokoyama, K. (Eds.), Computer
  Mathematics: Proceedings of the Fifth Asian Symposium. World Scientific
  Publishing Company, Singapore, pp. 165--174.

\bibitem[{Wang(2001{\natexlab{b}})}]{w01e}
Wang, D., 2001{\natexlab{b}}. {Elimination Methods}. Texts and Monographs in
  Symbolic Computation. Springer-Verlag, Wien New York.

\bibitem[{Wang(2004)}]{w04e}
Wang, D., 2004. {Elimination Practice: Software Tools and Applications}.
  Imperial College Press, London.

\bibitem[{Wu(1986)}]{w86b}
Wu, W.-T., 1986. Basic principles of mechanical theorem proving in elementary
  geometries. Journal of Automated Reasoning 2~(3), 221--252.

\bibitem[{Wu and Gao(2007)}]{WX2007M}
Wu, W.-T., Gao, X.-S., 2007. Mathematics mechanization and applications after
  thirty years. Frontiers of Computer Science in China 1, 1--8.

\bibitem[{Yang and Zhang(1994)}]{YZ1994S}
Yang, L., Zhang, J., 1994. Searching dependency between algebraic equations: an
  algorithm applied to automated reasoning. In: Johnson, J., McKee, S., Vella,
  A. (Eds.), Artificial Intelligence in Mathematics. Oxford University Press,
  Oxford, pp. 147--156.

\end{thebibliography}
% Include the ".bib" file (generated by bibtex) right here.

\section*{Appendix}
In this appendix we provide more details about the outputs of the
five algorithms in comparison, including degree tuples, numbers of
terms, head monomials, and maximal digits of coefficients of the
output polynomials, for some (but not all, due to space limitation)
of the examples.

Table \ref{tab:degree tuples} collects the degree tuples of the
polynomials in the output of each algorithm. Let us explain the
meanings of the entries in this table using an example: the entry in
the second row and the third column containing three tuples implies
that the output of $\algcs$ for example 2 consists of three
polynomials, say $F,G,H$ with $F<G<H$, the degrees of $F$ in
$w,x,y,z$ ($w<x<y<z$) are $520, 42, 0, 0$ respectively. One can see
that for most examples, the output polynomials of $\algcharsetw$ and
$\algcharset$ are of lower degrees than those of $\algcsw$ and
$\algcs$. The entry in the second row and the last column contains
$\ldots$, which means that there are more elements.

Statistical data about the numbers of terms of expanded polynomials
in the output of each algorithm are given in Table \ref{tab:nops2}.
The entry $6,6,6$ in the second column and the second row indicates
that for example 1, $\algcsw$ outputs a set of three polynomials,
which consists of $6,6,6$ terms respectively. For example 3 there
are four polynomials in $7$ variables. The output of any of the
first four algorithms contains four polynomials, and for
$\algcharset$ some of the polynomials are of degree $>100$. This may
explain why $\algcharset$ is slower than $\algcs$. Similar
situations occur for examples 8 and 16. For examples 2, 6, and 12,
the polynomials in the output of $\algcharset$ have fewer terms than
those in the output of $\algcs$, so $\algcharset$ is faster. For
most examples, the polynomials contained in the output of
$\algcharsetw$ have fewer terms, so $\algcharsetw$ is faster than
$\algcsw$, $\algcs$, and $\algcharset$. For examples 3, 14, and 15,
$\alggb$ outputs more polynomials of more terms than $\algcharset$.
It is therefore slower than $\algcharset$.

Table \ref{tab:head terms} displays the head terms of the polynomials in the
output of each algorithm. Because most of the initials are quite complicated,
we have reproduced only the head terms instead of the initials. Table
\ref{tab:digits} shows the maximal digits of coefficients of the polynomials
in the output of each algorithm. One sees clearly that the polynomials in the
outputs of $\algcharset$ and $\algcharsetw$ have smaller coefficients.

\begin{table}[]\footnotesize\caption{Degree tuples of polynomials in the outputs of five algorithms}
\centering
\begin{tabular}{cccccl}
\toprule
 No &$\algcsw$ & $\algcs$        & $\algcharsetw$ & $\algcharset$ & $\alggb$\\
%\hline
%DiscrC2     & $[2, 2, 1, 1, 1, 1, 1, 1, 0, 0, 0, 0]$ & $[2, 2, 1, 1, 1, 1, 1, 1, 0, 0, 0, 0]$ & $[2, 2, 1, 1, 1, 1, 1, 1, 0, 0, 0, 0]$  & $[2, 2, 1, 1, 1, 1, 1, 1, 0, 0, 0, 0]$ & \\
%            & $[2, 2, 1, 1, 0, 1, 0, 0, 1, 1, 1, 0]$ & $[2, 2, 1, 1, 0, 1, 0, 0, 1, 1, 1, 0]$ & $[2, 2, 1, 1, 0, 1, 0, 0, 1, 1, 1, 0]$  & $[2, 2, 1, 1, 0, 1, 0, 0, 1, 1, 1, 0]$ &  \\
%            & $[2, 2, 0, 1, 1, 0, 1, 0, 0, 1, 1, 1]$ & $[3, 3, 1, 1, 1, 1, 1, 0, 1, 1, 0, 1]$ & $[2, 2, 0, 1, 1, 0, 1, 0, 0, 1, 1, 1]$  & $[3, 3, 1, 1, 1, 1, 1, 0, 1, 1, 0, 1]$ &  \\
\midrule
2           &       & $[520, 42, 0, 0]$ & $[8, 12, 0, 0]$ &  $[8, 12, 0, 0]$ &  $[8, 12, 0, 0]$         \\ %Epsilon-A14
            &       & $[6, 30, 1, 0]$   & $[4, 6, 1, 0]$  &  $[4, 6, 1, 0]$  &  $[14, 11, 1, 0]$        \\
            &       & $[520, 40, 0, 1]$ & $[2, 1, 1, 1]$  &  $[4, 6, 0, 1]$  &  $[10, 10, 1, 0],\ldots$  \\
%\hline
%Trinks1     &     &     & $[2, 0, 0, 0, 0, 0]$ & $[2, 0, 0, 0, 0, 0]$ & $[2, 0, 0, 0, 0, 0]$ \\
%            &     &     & $[1, 1, 0, 0, 0, 0]$ & $[1, 1, 0, 0, 0, 0]$ & $[1, 1, 0, 0, 0, 0]$ \\
%            &     &     & $[1, 1, 1, 0, 0, 0]$ & $[1, 0, 1, 0, 0, 0]$ & $[1, 0, 1, 0, 0, 0]$ \\
%            &     &     & $[1, 1, 2, 1, 0, 0]$ & $[1, 0, 0, 1, 0, 0]$ & $[1, 0, 0, 1, 0, 0]$ \\
%            &     &     & $[1, 1, 2, 0, 1, 0]$ & $[1, 0, 0, 0, 1, 0]$ & $[1, 0, 0, 0, 1, 0]$ \\
%            &     &     & $[1, 0, 1, 0, 0, 1]$ & $[1, 0, 0, 0, 0, 1]$ & $[1, 0, 0, 0, 0, 1]$ \\
\hline
 5          & $[40, 0, 0, 0]$ & $[73, 0, 0, 0]$ & $[36, 0, 0, 0]$ & $[30, 0, 0, 0]$ & $[24, 0, 0, 0]$ \\ %ZeroDim14
            & $[39, 1, 0, 0]$ & $[53, 1, 0, 0]$ & $[25, 1, 0, 0]$ & $[25, 1, 0, 0]$ & $[23, 1, 0, 0]$ \\
            & $[39, 0, 1, 0]$ & $[70, 0, 1, 0]$ & $[3, 2, 1, 0]$  & $[29, 0, 1, 0]$ & $[23, 0, 1, 0]$  \\
            & $[1, 1, 1, 1] $ & $[71, 0, 0, 1]$ & $[1, 1, 1, 1]$  & $[29, 0, 0, 1]$ & $[23, 0, 0, 1]$  \\
\hline
 6          & $[146, 0, 0, 0]$ & $[168, 0, 0, 0]$ & $[28, 0, 0, 0]$ & $[27, 0, 0, 0]$ & $[24, 0, 0, 0]$ \\%Epsilon-A16
            & $[144, 1, 0, 0]$ & $[146, 1, 0, 0]$ & $[26, 1, 0, 0]$ & $[26, 1, 0, 0]$ & $[23, 1, 0, 0]$ \\
            & $[145, 0, 1, 0]$ & $[167, 0, 1, 0]$ & $[1, 3, 1, 0]$  & $[26, 0, 1, 0]$ & $[23, 0, 1, 0]$ \\
            & $[145, 0, 0, 1]$ & $[166, 0, 0, 1]$ & $[0, 2, 0, 1]$  & $[26, 0, 0, 1]$ & $[23, 0, 0, 1]$ \\
%\hline
%Epsilon-A25 & $[4, 5, 4, 6, 9, 0, 0, 0]$ & $[4, 4, 4, 5, 9, 0, 0, 0]$& $[8, 6, 6, 7, 12, 0, 0, 0]$& $[20, 11, 15, 10, 10, 0, 0, 0]$ & $[4, 4, 4, 5, 8, 0, 0, 0]$ \\
%            & $[2, 2, 2, 2, 5, 1, 0, 0]$ & $[2, 2, 2, 2, 5, 1, 0, 0]$& $[5, 4, 4, 4, 7, 1, 0, 0]$ & $[5, 4, 4, 4, 7, 1, 0, 0]$      & $[5, 5, 6, 6, 7, 1, 0, 0]$ \\
%            & $[2, 2, 2, 3, 6, 0, 1, 0]$ & $[2, 2, 2, 3, 6, 0, 1, 0]$& $[1, 1, 1, 1, 1, 1, 1, 0]$ & $[5, 4, 4, 5, 8, 0, 1, 0]$      & $[4, 4, 4, 5, 6, 1, 0, 0]$ \\
%            & $[0, 0, 0, 1, 1, 1, 1, 1]$ & $[2, 2, 2, 3, 6, 0, 0, 1]$& $[1, 0, 0, 0, 1, 1, 1, 1]$ & $[25, 16, 20, 15, 9, 0, 0, 1]$  & $[4, 4, 5, 5, 7, 1, 0, 0],\ldots$ \\
\hline
12          &  & $[675, 0, 0]$ & $[112, 0, 0]$ & $[81, 0, 0]$ & $[54, 0, 0]$ \\ %Steidel2
            &  & $[608, 1, 0]$ & $[102, 1, 0]$ & $[80, 1, 0]$ & $[53, 1, 0]$ \\
            &  & $[674, 0, 1]$ & $[4, 7, 1]$   & $[80, 0, 1]$ & $[53, 0, 1]$ \\
\hline
15          & $[153, 0, 0]$ & $[158, 0, 0]$ & $[161, 0, 0]$ & $[161, 0, 0]$ & $[141, 0, 0]$ \\ %Sym3-5
            & $[120, 1, 0]$ & $[120, 1, 0]$ & $[128, 1, 0]$ & $[128, 1, 0]$ & $[140, 1, 0]$ \\
            & $[1, 5, 1]$   & $[125, 0, 1]$ & $[5, 1, 1]$   & $[133, 0, 1]$ & $[140, 0, 1]$ \\
\hline
16          & $[16, 0, 0, 0]$ & $[14, 0, 0, 0]$ & $[16, 0, 0, 0]$ & $[14, 0, 0, 0]$ & \\ %Epsilon-A3
            & $[8, 1, 0, 0]$  & $[8, 1, 0, 0]$  & $[8, 1, 0, 0]$  & $[8, 1, 0, 0]$  & \\
            & $[2, 1, 1, 0]$  & $[9, 0, 1, 0]$  & $[2, 1, 1, 0]$  & $[9, 0, 1, 0]$  & \\
            & $[9, 0, 0, 1]$  & $[9, 0, 0, 1]$  & $[1, 1, 1, 1]$  & $[9, 0, 0, 1]$  & \\
%\hline
%17          & & & $[16, 0, 0, 0]$ & &   \\ %Wu90
%            & & & $[8, 1, 0, 0]$  & &   \\
%            & & & $[2, 1, 1, 0]$  & &   \\
%            & & & $[1, 1, 1, 1]$  & &   \\

\bottomrule
\end{tabular}\label{tab:degree tuples}
 \end{table}

\begin{table}[]\footnotesize\caption{Numbers of terms of polynomials in the outputs of five algorithms}
\centering
\begin{tabular}{cccccl}
\toprule
 No        &$\algcsw$ & $\algcs$  & $\algcharsetw$ & $\algcharset$ & $\alggb$\\

\midrule
  1   & $6,6,6$  & $6,6,8$  &  $6,6,6$ & $6,6,8$ &   \\  %DiscrC2
\hline
 2    &          & $5450,93,10390$ & $23,12,5$ &  $23,12,17$  & $23,45,35,39,5,$ \\  %Epsilon-A14
            &          &                 &           &              & $30,31,41,3$\\
\hline
3  & $18,26,38,2$ & $18,26,38,33$ & $18,26,5,2$ & $18,26,114,108$ & $10,8,5,17,25,9,\ldots$  \\ %Chou156-1
\hline
4   &   &   & $3,4,4,6,6,4$  & $3,4,4,4,4,4$  & $3,3,3,3,3,3$  \\  %Trinks1
\hline
5    & $41,80,80,4$  & $74,107,142,143$  & $37,51,12,4$  & $37,51,60,60$  & $25,25,25,25$\\ %ZeroDim14
\hline
6 & $143,284,284,286$ & $165,288,327,328$ & $25,50,8,5$ & $25,50,48,50$ & $25,25,25,25$  \\ %Epsilon-A16
\hline
7        & $292,522$ & $292,522$ & $88,158$ & $88,158$ & $88,393,498$ \\ %Schiele1
\hline
8 & $128,21,35,4$ & $89,21,35,35$ & $253,116,7,4$ & $3217,116,152,15087$ & $89,220,119,150,\ldots$ \\    %81,50,47,32,21,19,4,32,15,12,7,19,11,17,4,6,6  Epsilon-A25
\hline
  9         &  &  & $28,55,8,10,5$ &  & $4,8,10,5,12,13,$ \\%Cycli5
            &  &  &                &  & $5,12,15,13,5$\\
\hline
10        & $57,112,112,112$  & $103,153,180,204$  & $38,137,45,6$  &$60,118,118,118$  & $25,26,27,25,27$  \\ %Fee1
\hline
11  &    & $445,792,887$ & $70,140,8$ & $54,140,41$ & $54,55,55$ \\%Weispfenning94
\hline
 12   &  & $663,1188,1324$  & $108,198,13$ & $81,162,162$ & $53,55,55$  \\ %Steidel2
\hline
13      &   &      & $3752,29,29,12,$   &     & $172,11,10,11,11,$ \\ %Epsilon-A30
            &   &      & $8,6,8,6,7,6$   &      & $10,11,11,11,10$ \\
\hline
14     &   &      & $87,199,16$    & $76,152,150$ & $45,48,48,48,\ldots$ \\ %Fateman
\hline
 15   & $139,153,3$ & $139,153,152$ & $90,88,3$ & $90,88,88$ & $90,142,142$ \\ %Sym3-5
\hline
16 & $3123,210,8,547$ & $1984,210,378,547$ & $20933,210,8,6$ & $8169,210,378,547$ &  \\ %Epsilon-A3
\hline
 17       &  &  & $49226,305,8,6$ &  &  \\ %Wu90
\bottomrule
\end{tabular}\label{tab:nops2}
 \end{table}
%\note{Wang: Please try to order the submonomials uniformly in the last table.}
\begin{table}[]\footnotesize\caption{Head terms of polynomials in the outputs of five algorithms}
\centering
\begin{tabular}{cccccc}
\toprule
 No         &$\algcsw$ & $\algcs$        & $\algcharsetw$ & $\algcharset$ & $\alggb$\\
\midrule%DiscrC2
 1    & $f,y^2 i,y^2 j$  & $f,y^2 i,y^3 j$  & $f,y^2 i,y^2 j$   & $f,y^2 i,y^3 j$  &   \\

\hline%Epsilon-A14
2 &       & $w^{508}x^{42},{x^{29}y},$ &$x^{12},{w^2x^3y},$& $x^{12},{w^2x^3y}  ,$ &$x^{12},w^{14}y,$   \\
  &       & $w^{510}x^{40}z$             &$xyz$                & $x^6 z$             &${w^{10}xy},\ldots$   \\
\hline%ZeroDim14
5        &$x_1^{40},x_1^{39} x_2,$ &$x_1^{73},x_1^{52} x_2,$   &$x_1^{36},{x_1^{24}x_2},$  &$x_1^{30},{x_1^{24}x_2}, $ & $x_1^{24},x_2,x_3,x_4$\\
            &$x_1^{39} x_3,x_4$       &$x_1^{70} x_3,x_1^{70} x_4$&${x_2^2x_3},x_4$           &$x_1^{29}x_3,x_1^{29}x_4$ &    \\
\hline %Epsilon-A16
6 & $d^{146},d^{144}p,$ & $d^{168},d^{146}p,$ &$d^{28},d^{26}p,$ & $d^{27},d^{26}p,$ & $d^{24},p,c,q$ \\
            & $d^{145}c,d^{145}q$ & $d^{166}c,d^{166}q$ &$p^3c,pq$         & $d^{26}c,d^{26}q$ &    \\
\hline %Epsilon-A25
8 &$t^9d^2,t^5z,$&$t^9d,t^5z, $&$t^{12} d a^2,t^6 d az,$ &$t^{10}a^3 b^2 c^6 d^5 ,t^6adz ,$   &$t^8 d,c^2 d^4 z,$ \\
  &$t^6y,tx$     &$t^6 y,t^6 x$&$zy ,yx $          &$t^7 a d y,t^8 a^3 b^2 c^6 d^{11}x$ & $tc^2 d^2 z , t^2 bc^3 z,\ldots$  \\

\hline %Fee1
10        & $q^{56},q^{55}c,$ & $q^{104},q^{78} c,$   & $q^{73},q^{62}c,$ &$q^{60},q^{59}c,$ & $q^{24},qc,$   \\
          & $q^{55}p,q^{55}d$ & $q^{91}p,q^{103}d $ &   $q^4c^2p,qpd$  &$q^{59}p,q^{59}d$ & $c^3,p,d$   \\

\hline %Steidel2
12          &  & $x^{675},x^{608}y,$ &$x^{112},x^{102}y,$  &  $x^{81},x^{80}y,$ & $x^{54},y,z$    \\
            &  & $x^{674}z$         &$x^3y^5z$           &  $x^{80}z$        &            \\

\hline %Sym3-5
15          &  $x^{153},x^{120}y,z$ &  $x^{158},x^{120}y,$ & $x^{161},x^{128}y,$ &$x^{161},x^{128}y,$ & $x^{141},y,z$\\
            &                       &    $x^{125}z$       & $x^5z$             & $x^{133}z$        &              \\

\hline %Epsilon-A3
16   & $x_{71}^{16},x_{71}^7x_{72},$ &$x_{71}^{14},x_{71}^7x_{72},$ & $x_{71}^{16},x_{71}^7x_{72},$ & $x_{71}^{14},x_{71}^7x_{72},$  & \\
            & $x_{71}^2x_{73},x_{71}^9x_{80}$ & $x_{71}^9x_{73},x_{71}^9x_{80}$ & $x_{71}^2x_{73},x_{71}x_{73}x_{80}$& $x_{71}^9x_{73},x_{71}^9x_{80}$&  \\

\hline%Wu90
17        &  &  & $x_1^{16},x_1^7x_2,$ &  &  \\
             &  &  & $x_1^2x_3,x_1x_3x_4$ &  &  \\
\bottomrule
\end{tabular}\label{tab:head terms}
\end{table}

\begin{table}[]\footnotesize\caption{Maximal numbers of digits of coefficients of polynomials in the outputs of five algorithms}
\centering
\begin{tabular}{cccccc}
    \toprule
    No                &$\algcsw$          & $\algcs$          & $\algcharsetw$& $\algcharset$     & $\alggb$     \\
    \midrule
     1         &$1,1,1$            & $1,1,1$           & $1,1,1$       & $1,1,1$           &    \\\hline
     2     &                   & $212,3,212$       &  $2,1,1$      & $2,1,1$           & $2,3,3,\ldots$   \\\hline
     3       & $2,2,2,1$         & $2,2,2,2$         & $2,2,1,1$     & $2,2,4,4$         & $1,1,1,\ldots$   \\\hline
     4         &                   &                   &$5,15,5,7,7,3$ &$5,15,14,24,23,13$ &$5,5,4,5,4,3$    \\\hline
     5       & $40,110,146,1$    & $71,55,70,70$     & $20,27,4,1$   & $20,28,215,216$   & $20,257,258,245$   \\\hline
     6     & $158,164,776,468$ & $184,164,188,189$ & $39,36,2,1$   & $18,22,614,298$   & $18,266,271,266$   \\\hline
     7        & $11,11$     & $11,11$   & $6,7$  & $6,6$  & $6,15,17$   \\\hline
     8     & $2,2,2,1$ & $2,2,2,2$   & $5,4,1,1$ & $6,2,2,9$  & $2,3,3,\ldots$   \\\hline
     9         &      &    & $14,21,1,1,1$  &   & $3,3,6,\ldots$  \\\hline
     10            & $36,123,330,383$ & $67,50,60,288$ & $20,160,9,1$  & $72,264,603,904$  & $15,163,186,143,182$   \\\hline
     11  &      & $94,85,97$   & $18,17,1$  & $8,15,298$  & $8,202,202$   \\\hline
     12        &      & $89,78,99$   & $16,14,1$  & $12,22,364$  & $9,193,194$   \\\hline
     13     &      &    & $3,1,1,1,1,1,\ldots$  &   & $2,1,1,1,1,1,\ldots$   \\\hline
     14         &      &    & $17,19,1$  & $16,64,686$  &  $10,174,170,\ldots$  \\\hline
     15          & $13,11,1$ & $13,11,11$ & $9,7,1$  & $9,7,7$  & $9,625,626$   \\\hline
     16      & $6,3,1,3$ & $5,3,3,3$ &  $8,3,1,1$ & $6,3,3,3$   &    \\\hline
     17            &      &    & $8,3,1,1$ &   &    \\\hline
     18         &      &    &   &   &   $6,17,19,\ldots$ \\
\bottomrule
\end{tabular}\label{tab:digits}
 \end{table}
\end{CJK*}
\end{document}